\documentclass[a4paper,11pt]{article}

\usepackage{lmodern}
\usepackage{jcappub} 
\usepackage{packages,bm}
\usepackage{hyperref}

\def\iMpch{\,h\,{\rm Mpc}^{-1}}
\def\reffig#1{Fig.~\ref{fig:#1}}
\def\refeq#1{Eq.~\eqref{eq:#1}}
\def\refeqs#1#2{Eqs.~\eqref{eq:#1}--\eqref{eq:#2}}

\title{Mixed Dark Matter and Galaxy Clustering: The Importance of Relative Perturbations}

\author[a]{\c{S}afak \c{C}elik}
\author[a]{and Fabian Schmidt}

\affiliation[a]{Max Planck Institute for Astrophysics,
Karl–Schwarzschild–Straße 1, D–85748 Garching,
Germany}

\emailAdd{scelik@mpa-garching.mpg.de}
\emailAdd{fabians@mpa-garching.mpg.de}

\abstract{We develop a perturbative model to describe large-scale structure in cosmologies where dark matter consists of a mixture of cold (CDM) and warm (WDM) components. In such mixed dark matter (MDM) scenarios, even a subdominant warm component can introduce distinctive signatures via its free-streaming effects, altering the evolution of density and velocity perturbations. We present linear-order solutions for both total and relative perturbations in the two-fluid system, identifying novel contributions to galaxy bias caused by the relative density and velocity modes between the components. Incorporating these effects into the galaxy bias expansion, we compute the linear galaxy power spectrum in both real and redshift space. Using Fisher matrix forecasts, we assess the sensitivity of upcoming surveys such as DESI and PFS to MDM scenarios. Our results demonstrate that neglecting relative perturbations can lead to significant biases in inferred constraints on the warm dark matter fraction, particularly for lighter WDM masses ($\lesssim 150~\mathrm{eV}$ and $\lesssim 80~\mathrm{eV}$) for PFS and DESI, respectively. This framework provides a consistent and generalizable approach for incorporating multi-component dark matter dynamics into galaxy clustering analyses.
}

\begin{document}
\maketitle

\section{Introduction}
The $\Lambda$CDM model is the most widely accepted framework for describing the major components of the Universe. It posits the existence of a cosmological constant ($\Lambda$), responsible for the observed late-time acceleration of the Universe, and a dark matter component that is predominantly ``cold''---meaning it has a negligible primordial velocity dispersion. Despite its simplicity, $\Lambda$CDM has been remarkably successful in explaining a wide range of cosmological observables, including the anisotropies of the Cosmic Microwave Background (CMB), which probe the early Universe, and the distribution of Large-Scale Structures (LSS), which trace the growth of structure at later times.

However, the increasing precision of cosmological observations has revealed several tensions that suggest $\Lambda$CDM may be an incomplete description of our Universe. For instance, the cold dark matter (CDM) assumption has been challenged by various observations probing different scales in the universe\cite{Bullock_2017,Weinberg2015}. On galactic scales, CDM simulations predict cuspy central density profiles in dark matter halos, whereas observations of dwarf and low-surface-brightness galaxies often favor cored profiles—a discrepancy known as the ``core-cusp problem.'' \cite{deBlok2010,Moore1994} CDM also over-predicts the abundance of satellite galaxies around Milky Way–like halos, commonly referred to as the ``missing satellites problem'' \cite{Klypin1999,Moore1999}. Furthermore, Lyman-$\alpha$ forest data and weak lensing measurements from surveys such as KiDS and DES suggest a possible suppression of small-scale power relative to $\Lambda$CDM predictions \cite{Asgari_2021,Abbott2022}.

There have been several theoretical and phenomenological studies on different alternatives of dark matter to overcome these possible issues with CDM. One class of alternatives are Axion like particles (ALP's) that erase small-scale structures due to the wave-like behaviour that they have on small scales \cite{Ferreira_2021}. Another is warm dark matter (WDM), consisting of particles with non-negligible velocity dispersion that suppresses structure formation below a characteristic scale. The latter include various different theoretical models including thermal freeze-out \cite{KolbTurner1990}, non-resonant (Dodelson-Widrow) \cite{DodelsonWidrow1994} or resonant (Shi-Fuller) \cite{ShiFuller1999} sterile neutrino production, or decay of heavier particles \cite{Fuss:2022zyt} which also suppress small scale structure growth due to their non-negligible velocity dispersion. 

Warm dark matter (WDM) models are tightly constrained by a range of cosmological and astrophysical observations. Some of the main methods include the Lyman-$\alpha$ forest power spectrum \cite{Viel:2013fqw}, which probes small-scale structure at high redshift, satellite galaxy counts \cite{DES:2019ltu}, and strong lensing substructure analyses \cite{Enzi_2021}. These probes are sensitive to the free-streaming effects of WDM, which suppress the formation of small-scale structure, thereby placing lower bounds on the WDM particle mass.

However, almost all constraints using the above methods assume that WDM makes up all of the dark matter. If WDM constitutes only a sub-dominant portion of the total dark matter density, many of these constraints become significantly relaxed. In mixed dark matter (MDM) models, a dominant cold component drives efficient structure formation, while a subdominant warm component subtly modifies growth, especially on intermediate and mildly nonlinear scales. Consequently, WDM particle masses that would be ruled out in pure-WDM scenarios can remain viable when a dominant cold component is present. Allowing a subdominant non-cold component can also introduce parameter degeneracies with other sectors; for example, \cite{Stadler:2018dsa} showed that MDM with a photon-interacting subcomponent can mimic the small-scale suppression from massive neutrinos in large-scale structure, complicating neutrino-mass inference in DESI-like analyses.

In this work, we investigate the phenomenology of MDM scenarios, where the dark matter sector contains both cold and warm components. Specifically, we aim to understand how a subdominant warm component, such as a sterile neutrino produced via the Dodelson-Widrow mechanism, affects large-scale structure observables. Our primary focus is on large-scale galaxy clustering, with the linear galaxy power spectrum in both real and redshift space serving as the main probe. This approach can be generalized to other cosmological observables such as Lyman-$\alpha$ forest power spectrum by using the selection dependent bias formalism developed in \cite{Desjacques:2018pfv,Ivanov:2023yla}. 

MDM scenarios have been already explored as a way to reconcile small-scale tensions in $\Lambda$CDM in the literature. Several studies have investigated cosmological constraints on such models using different observables. Early work by~\cite{Boyarsky_2009} used Lyman-$\alpha$ forest data to place stringent limits on the warm fraction and mass, demonstrating that even a small warm component can significantly alter small-scale structure. The implementation of MDM scenarios in the Boltzmann solver \texttt{CLASS}~\cite{Lesgourgues_2011,lesgourgues2011cosmiclinearanisotropysolving} enabled more recent analyses, such as~\cite{Xu:2021rwg, Peters:2023asu}, who estimated non-linear structure formation in MDM models either with an analytic method or N-body simulations, and performed joint analyses with CMB, large-scale structure data and weak lensing measurements. Refs.~\cite{Xu:2021rwg, Peters:2023asu} provided updated bounds on the parameter space of thermal relics and sterile neutrino dark matter respectively. These studies highlight the importance of combining multiple observables to constrain MDM scenarios. However, they have so far typically relied on single-fluid treatments and do not account for the dynamical effects of relative perturbations between the species. Complementary non-linear treatments include \cite{Parimbelli:2021mtp}, who built an emulator for MDM to predict the matter power spectrum and examined impacts on weak lensing, galaxy clustering, and the halo mass function, and \cite{Dome:2024hzq}, who calibrated a halo-model framework for mixed axion–CDM cosmologies against dedicated simulations.

A key feature of our approach is the inclusion of relative perturbations between the cold and warm components and their impact on galaxy bias. By \textit{relative perturbations}, we refer to the minimal relative density and velocity perturbations arising from the initial difference in velocity dispersion between the two components, which naturally emerge due to the free-streaming of WDM. More generally, for example in the case of initial conditions generated by multi-field inflation, primordial iso-curvature perturbations between the components could also be present, potentially leaving additional imprints on large-scale structure; see \cite{Barreira:2023uvp} for a corresponding study of primordial baryon-CDM iso-curvature perturbations. Although we do not consider these in this work, they could be added straightforwardly.

We focus on the general bias expansion within the effective field theory of large-scale structure (EFTofLSS), which has been widely used for single-species dark matter. Within the EFT, the presence of multiple species with distinct perturbation histories introduces additional bias operators \cite{Schmidt_2016}.
Throughout, we restrict our analysis to scales larger than the WDM free-streaming scale, $\lambda_{\rm fs}$, because extending the EFT to scales at or below $\lambda_{\rm fs}$ requires additional assumptions and/or an expansion in the WDM mass fraction \cite{Verdiani:2025jcf, Senatore:2017hyk}. This choice confines our treatment to WDM particle masses $\gtrsim 10\,\mathrm{eV}$, without requiring the WDM mass fraction to be much less than unity. As such, our approach is complementary to Ref.~\cite{Verdiani:2025jcf}.

The paper is structured as follows: In Section~\ref{sec:linear matter}, we present the linear evolution of matter perturbations in the presence of both cold and warm dark matter, including a treatment of relative modes. In Section~\ref{sec:linear galaxy bias}, we construct the linear galaxy power spectrum in both real and redshift space, accounting for the modified bias expansion. Section~\ref{sec:Fisher} presents our Fisher forecast analysis, where we assess the constraining power of DESI \cite{DESI:2016fyo} and  the Prime Focus Spectrograph (PFS) survey on MDM models with different warm component masses and fractions. We conclude in Section~\ref{sec:Conclusion} with a summary and discussion of future directions.

\section{Linear-Order Solutions for the Matter Field} \label{sec:linear matter}
\subsection{WDM and Structure Formation}
In this section, we review the description of WDM in the context of cosmological structure formation. First, it is important to emphasize that any constraint on WDM must be interpreted in the context of its production mechanism. Ultimately, observational constraints, particularly those from large-scale structure (LSS), do not directly probe the particle mass itself, but rather the suppression of power in the matter power spectrum caused by the free-streaming of the WDM component. This suppression depends not only on the mass, but also on the particle's momentum distribution, which is determined by how it was produced. While we have referred to the scale $\lambda_\mathrm{fs}$, it is worth pausing to clarify its physical meaning. The free-streaming length  at a given time is the typical distance that a warm particle travels within a Hubble time, and is defined as:
\begin{equation}
   k_\mathrm{fs}^{-1}(t) = \frac{\lambda_{\mathrm{fs}}(t)}{2\pi} \sim \frac{\langle v^2(t)\rangle^{1/2} }{\mathcal{H}(t)},
\end{equation}
where $\langle v^2(t)\rangle^{1/2}$ is the RMS WDM particle velocity dispersion, obtained by averaging over the phase-space distribution function, and $\mathcal{H}(t) = a H(t)$ is the comoving Hubble rate. This provides an estimate of the free-streaming scale, $\lambda_{\rm fs}$, below which no particle can gravitationally cluster—random motions erase perturbations on all smaller scales.

For example, consider two WDM candidates with the same mass of 100 eV but produced via different mechanisms. A thermal relic that decoupled from the plasma before Standard Model (SM) neutrinos would miss the subsequent entropy injections; see, e.g., Ref.~\cite{Xu:2021rwg}, which presents a simple approximate relation between the relic temperature of a thermal WDM particle and its mass. As a result, its relic temperature today would be lower than that of the SM neutrinos, leading to a shorter free-streaming length and a milder suppression of small-scale structure. It also matters whether the relic is fermionic or bosonic, since Fermi–Dirac versus Bose–Einstein statistics change the relation between mass, temperature, and abundance, and thus the free-streaming scale. In contrast, a sterile neutrino produced via a non-thermal mechanism, such as the Dodelson-Widrow scenario, would share a relic temperature similar to that of SM neutrinos. Despite having the same mass (100 eV), its momentum distribution would be warmer, resulting in a significantly larger free-streaming scale and a stronger suppression of small-scale power.

This illustrates that mass bounds derived from LSS observables cannot be universally applied without specifying the underlying production model. The mapping between mass and free-streaming length is model-dependent, and failing to account for this can lead to misleading conclusions about the viability of a given WDM scenario.

This example highlights why non-resonantly produced sterile neutrinos are particularly convenient for our purposes. 
Unlike thermally produced warm particles---whose relic temperature depends on the specific decoupling history and must be redefined for each mass---sterile neutrinos generated through non-thermal mechanisms typically exhibit a relic temperature close to that of active neutrinos, regardless of their mass. For such particles we have a simple relation in between $k_{\rm fs}$ and $m_w$ as \cite{shoji2010massive}:
\begin{align} \label{eq:free streaming def}
    k_{\rm fs}(z) = \frac{0.677}{(1+z)^2}\left(\frac{m_w}{1\; \rm eV}\right)[\Omega_m\;(1+z)^3]^{1/2} h \; \mathrm{Mpc}^{-1} 
\end{align}

Moreover, their initial momentum distribution is well approximated by a Fermi--Dirac function rescaled by a normalization factor~\cite{DodelsonWidrow1994},
\begin{equation}
    f(p,z) = \frac{A}{e^{p/T_{\nu}(z)} + 1},
\end{equation}
where $A$ is determined by the mean energy density of the warm component given its mass, and $T_\nu$ denotes the neutrino temperature today. 
It is convenient to express this energy density in terms of the \emph{fraction of the warm component},
\begin{equation}
    f_w \equiv \frac{\Omega_w}{\Omega_c + \Omega_w} = \frac{\Omega_w}{\Omega_{\mathrm{DM}}},
\end{equation}
where $\Omega_w$ and $\Omega_c$ are the present-day density fractions of the warm and cold dark matter components, respectively.
Notice that $\Omega_w = f_w \Omega_{\rm DM}$. 
In the sterile-neutrino case, specifying the particle mass $m$ and the fraction $f_w$ completely fixes the normalization $A$, making the model fully determined by these two parameters. 
The \texttt{CLASS} Boltzmann solver\cite{lesgourgues2011cosmiclinearanisotropysolving} includes this class of particles in its default configuration, allowing users to specify just the relic temperature, particle mass, and energy density (or $f_w$); the code then automatically computes the appropriate normalization for the distribution function. 

\begin{figure}
    \centering
    \includegraphics[width=0.8\linewidth]{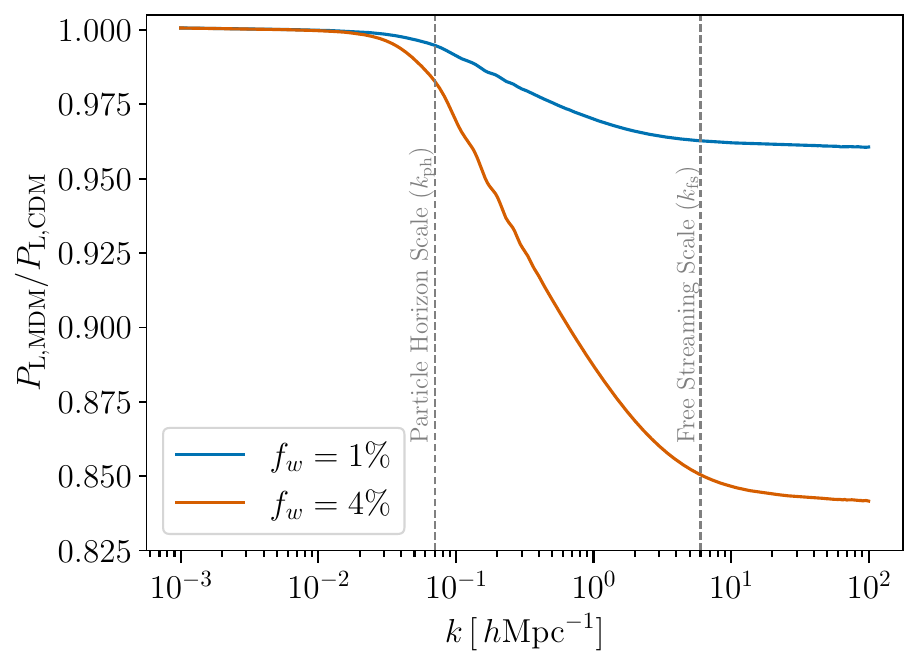}
     \caption{Characteristic scales of the linear matter power spectrum at $z=2$ in the presence of a mixture of cold and warm dark matter components, the latter having  $m_w = 10\:\mathrm{eV}$.}
    \label{fig:characteristic scales}
\end{figure}

In describing the impact of a warm component on the linear matter power spectrum $P_{\mathrm{lin}}$, it is useful to think in terms of \emph{two parameters}---the mass $m_w$ and fraction $f_w$---and \emph{two characteristic scales}---the free-streaming scale $k_{\rm fs}$ and the particle-horizon scale $k_{\rm ph}$. 
The particle-horizon scale quantifies the largest distance warm particles can travel over the age of the Universe and is given by
\begin{equation}
    k_{\mathrm{ph}}^{-1} \equiv \int_0^{t_\mathrm{obs}} \frac{\langle v^2(t)\rangle^{1/2}}{a(t)} \, dt,
\end{equation}
where $a(t)$ is the scale factor and the integral runs from the early Universe to the observation time $t_\mathrm{obs}$. 
Figure~\ref{fig:characteristic scales} illustrates the relevant scales and their roles in shaping the linear matter power spectrum, indicating that the suppression starts to emerge at $k_{\mathrm{ph}}$, and reaches its maximum at $k_{\mathrm{fs}}$, where its amplitude is governed by $f_w$. 

Even though the warm component initially has a significantly higher velocity dispersion compared to CDM, its temperature $T_\nu$ decays as $1/a$ as the universe expands. By the time it cools sufficiently (around $z\approx 30$), the evolution of perturbations for each component can be described using the fluid equations---the first two moments of the Boltzmann equation.
The fluid approximation is valid on scales larger than the free-streaming scale, $k \ll k_{\rm fs}$, at least for thermal relics \cite{shoji2010massive}, which is the regime we will focus on in this paper; we will consequently also neglect the small-scale Poisson-like fluctuations that can be present for warm components \cite{amin/etal:2025}. Depending on the WDM mass, this puts an upper limit on the range of scales that we can describe, in addition to the nonlinear scale $k_{\rm nl}$ that limits the fluid approximation for the CDM component. Note that we do not assume that $f_w$ is much less than unity.
We next derive the fluid equations describing the joint evolution of the CDM and WDM components.

\subsection{Fluid Equations for Warm \& Cold Dark Matter}
The evolution of collisionless dark matter particles is governed by the Vlasov equation (also known as the collisionless Boltzmann equation), which follows from Liouville's theorem applied to particles following Hamiltonian trajectories in phase space. This equation describes the time evolution of the phase space distribution function $f(\mathbf{x}, \mathbf{p}, \eta)$ for non-relativistic particles in comoving-Newtonian gauge:
\begin{equation}\label{eq:Vlasov}
    \frac{\partial f}{\partial \eta} + \frac{\mathbf{p}}{a m} \cdot \nabla_{\mathbf{x}} f - a m \nabla_{\mathbf{x}} \Phi \cdot \nabla_{\mathbf{p}} f = 0,
\end{equation}
where $\Phi(\mathbf{x}, \eta)$ is the Newtonian gravitational potential, $\mathbf{p}$ and $\mathbf{x}$ are the comoving momentum and comoving coordinates respectively, $\eta$ being the conformal time and $m$ is the mass of the dark matter particle. This form of the Vlasov equation applies to a \emph{single} dark matter species, such as cold or warm dark matter considered individually. In scenarios with multiple components, each having distinct phase space distribution functions (e.g., different temperatures or momentum distributions), the equation must be applied separately to each component rather than to the total distribution.
In this subsection, we consider equations for a single component (warm or cold), but refrain from adding a additional subscripts for clarity.
The fluid approximation proceeds by taking
moments of Eq.~\eqref{eq:Vlasov}. We define the moments of the Vlasov equation over momentum space as follows:
\begin{itemize}
    \item \textbf{Density:}
    \begin{equation}
        \rho = m \int d^3p \, f(\mathbf{x}, \mathbf{p}, \eta),
    \end{equation}

    \item \textbf{Mean (bulk) velocity:}
    \begin{equation}
        \rho v_i = \int d^3p \, \frac{p_i}{a} f ,
    \end{equation}

    \item \textbf{Velocity dispersion tensor:}
    \begin{equation}\label{eq:vel disp}
     \sigma_{ij}  + \rho v_iv_j= \int d^3p \, \frac{p_ip_j}{a^2m}  f,
    \end{equation}
    where $\mathbf{v} = \mathbf{p} / (a m)$ is the comoving peculiar velocity.
\end{itemize}
If one takes the zeroth moment of the Vlasov equation one obtains the continuity equation: 
\begin{equation}\label{eq:continuity}
    \frac{\partial \delta}{\partial \eta} + \bm{\nabla}\cdot[(1+\delta)\bm{u}] = 0,
\end{equation}
where we have introduced the density contrast for the component considered as $\delta = \delta\rho / \bar{\rho}$. Likewise, taking the first moment of the Vlasov equation, and contracting with $\partial_i$, yields the divergence of Euler's equation:
\begin{equation}\label{eq:euler's}
   \frac{\partial \theta}{\partial \eta} + \mathcal{H} \theta + \bm{\nabla}\cdot(\bm{u}\cdot\bm{\nabla}\bm{u}) = -\mathbf{\nabla}^2 \Phi - \frac{1}{\rho} \mathbf{\nabla}^2 (\sigma_{ij}).
\end{equation}
where we have defined the velocity divergence as $\theta = \nabla \cdot \mathbf{v}$. We have neglected the vorticity component of the velocity field, as it decays with time at any order in perturbation theory, and only becomes relevant on fully nonlinear scales. The velocity dispersion tensor $\sigma_{ij}$ can be decomposed into an isotropic (diagonal) and anisotropic (traceless) part:
\begin{equation}
    \sigma_{ij} = \mathcal{P}\delta_{ij} + \pi_{ij},
\end{equation}
where $\mathcal{P}$ is the effective pressure, and $\pi_{ij}$ is the anisotropic stress tensor, which accounts for the viscous behavior of the fluid. The latter is typically small and therefore often neglected; however, we retain the pressure term when considering the warm component. Substituting this into Eq.~\eqref{eq:euler's}, we obtain the divergence of the Euler equation with pressure:
\begin{equation}
    \frac{\partial \theta}{\partial \eta} + \mathcal{H}\theta + \bm{\nabla}\cdot(\bm{u}\cdot\bm{\nabla}\bm{u}) = -\mathbf{\nabla}^2 \Phi - \frac{1}{\rho} \mathbf{\nabla}^2 \mathcal{P},
\end{equation}
where the gravitational potential is determined by the density contrast via the Poisson equation: 
\begin{equation}\label{eq:poisson eq}
   \nabla^2 \Phi = 4\pi G a^2 \bar{\rho} \delta_m.
\end{equation}
Here, $\delta_m = f_c\delta_c + f_w\delta_w$ denotes the total matter density contrast, defined as the weighted sum of the density perturbations of two components  where $f_s$ for $s \in \{c, w\}$ denotes the fractional contribution of each species to the total dark matter density.
To close the system of equations, an equation of state for the pressure $\mathcal{P}$ is needed. We introduce an effective sound speed $c_s^2$ as
\begin{equation}
    c_s^2 \equiv \frac{\partial \mathcal{P}}{\partial \rho}.
\end{equation}
This expression is analogous to the thermodynamic definition of the sound speed and characterizes the response of pressure to small density perturbations. For warm dark matter, the pressure does not arise from interparticle interactions but instead originates from the intrinsic velocity dispersion of the collisionless particles—determined by how they were generated in the early universe. At leading order, the effective sound speed $c_s$ decreases proportionally to $1/a$ as the Universe expands, which can be seen directly through Eq.~\eqref{eq:vel disp}. Consequently, the extent to which the pressure term suppresses structure formation depends on both the time of particle production and the initial velocity dispersion. These factors are closely tied to the underlying particle generation process discussed earlier. In the case of CDM particles we can neglect the initial velocity dispersion so that we do not need to include the pressure term; the effective pressure induced by nonlinear evolution of small-scale perturbations \cite{Carrasco:2012cv,Baumann:2010tm,Ivanov:2022mrd} is of the same order as nonlinear corrections, whose treatment we defer to future work.

\subsection{Solutions of the Fluid Equations for a Two-Fluid System}

Linearizing \refeqs{continuity}{euler's} for the two components, CDM and WDM, respectively, we obtain the following set of equations to describe the linear evolution of the two components in our model:
\begin{align}\label{eq:linear boltzmann}
    &\frac{\partial}{\partial \eta} \delta_{c} + \theta_{c} = 0  \nonumber \\
     &\frac{\partial}{\partial \eta} \theta_{c} + \mathcal{H}\theta_{c} + \frac{3}{2}\Omega_{m}(a)\mathcal{H}^2 \delta_{m} = 0  \nonumber \\
     &\frac{\partial}{\partial \eta} \delta_{w} + \theta_{w} =  0  \nonumber \\
    &\frac{\partial}{\partial \eta} \theta_{w} + \mathcal{H}\theta_{w} + \frac{3}{2}\Omega_{m}(a)\mathcal{H}^2 \delta_{m} - c_s^2(a)k^2\delta_{w}  = 0, 
\end{align}
where we have moved to Fourier space for convenience and reformulated Eq.~\eqref{eq:poisson eq} with the help of the Friedmann equations \cite{Dodelson_Schmidt_2021}. Since gravity only cares about the total stress-energy, the gravity sourc term  is associated with the total matter density fluctuations ($\delta_m$) for both fluids, which are hence coupled via gravity.

The evolution of two-fluid systems, such as the baryon--cold dark matter (CDM) mixture, has been studied extensively in the literature~\cite{2009ApJ...700..705S,2010PhRvD..82h3520T,2015JCAP...05..019L,Schmidt_2016}. In those cases, both components can be treated as pressureless perfect fluids (since the baryonic pressure is negligible on quasilinear scales), and no pressure term appears in the perturbation equations for the baryons. As a result, by switching from individual fluid perturbations to a basis of total matter and relative perturbations, the system of four coupled first-order differential equations can be reduced to two decoupled second-order equations.

In our case, however, this simplification cannot be applied directly due to the presence of a pressure term in the warm component. Despite this, it remains useful to adopt the same basis transformation. With appropriate approximations, the system can still be effectively described in terms of total and relative perturbations, even though the equations for relative and total perturbations remain partially coupled.

We now define our new basis as follows: the relative density and velocity divergence perturbation is given by $\delta_r := \delta_c - \delta_w$, and $\theta_r := \theta_c - \theta_w$. In the same way we defined the total matter density perturbations after Eq.~\eqref{eq:poisson eq} we define $\theta_m := f_c \theta_c + f_w \theta_w$. After switching to the new basis Eq.~\eqref{eq:linear boltzmann} becomes:
\begin{align}
&\delta_m' + \theta_m = 0 \nonumber\\
&\theta_m' + \mathcal{H}\theta_m + \frac{3}{2}\Omega_m\mathcal{H}^2\delta_m - \frac{c_{s_i}^2k^2}{a^2}(f_w\delta_m - \cancel{f_wf_c\delta_r})=0 \nonumber\\
&\delta_r' + \theta_r= 0 \nonumber\\
&\theta_r' + \mathcal{H}\theta_r + \frac{c_{s_i}^2k^2}{a^2}(\delta_m - \cancel{f_c\delta_r})=0.
\label{eq:first order odes}
\end{align}
where $c_{s_i}$ is a dimensionless quantity defined as $c_s(a) = \frac{c_{s_i}}{a}$ which depends only on $k_{fs}$, and therefore implicitly on $m_w$. 

In the full set of equations describing the evolution of total and relative perturbations in a two-fluid system with cold and warm components, the equations for $\delta_m$ and $\theta_m$ are coupled to those of $\delta_r$ and $\theta_r$ through the pressure gradient term. Specifically, the pressure depends on $\delta_w$, which can be expressed as a linear combination of $\delta_m$ and $\delta_r$. 
However, since the relative density perturbation $\delta_r$ is not sourced by gravity, it is not expected to grow significantly (we will confirm this below). In contrast, $\delta_m$ is directly sourced by the gravitational potential and dominates the clustering behavior. Therefore, as a physically motivated approximation, we neglect the subdominant contribution proportional to $\delta_r$ in the pressure term, as indicated above. This simplifies the system by decoupling the total and relative perturbations. Notice also that, following our assumption, $\delta_r$ is only sourced by free-streaming, and hence suppressed by $(k/k_{\rm fs})^2$ on large scales, so that the neglected contributions in \refeq{first order odes} scale as $\sim k^4 \delta_m$.

Adopting an Einstein-de Sitter (EdS) universe for the moment, the approximated form of the fluid equations immediately provide us with the following set of solutions:
\begin{align}\label{eq:total matter solns}
    \delta_m ^{(1)}(\bm{k},a) =\:& a\delta_m^{(0)}(\bm{k}) - \frac{R_p(\bm{k})f_w}{3H_0} + \mathcal{O}\left(\frac{k}{k_{\mathrm{fs}}}\right)^4 \nonumber \\
    \theta_m^{(1)} (\bm{k},a) =\:& -a^{\frac{1}{2}}H_0\delta_m^{(0)}(\bm{k}) + \mathcal{O}\left(\frac{k}{k_{\mathrm{fs}}}\right)^4 \nonumber \\
    \mbox{with}\qquad  R_p(\bm{k}) =\:& -\frac{2c_{s_i}^2k^2}{H_0}\delta_m^{(0)}(\bm{k}).
\end{align}
We fix the normalization of the growing mode by defining the amplitude of the adiabatic total-matter perturbation at an initial redshift $z_{\mathrm{ini}}=10$ (well into matter domination) as
$
\delta_m^{(0)}(\bm{k})\,\equiv\,\delta_{m,\mathrm{ad}}(\bm{k},z_{\mathrm{ini}})/D_+(z_{\mathrm{ini}})\,,
$
where $D_+(z)$ is the linear growth factor normalized to $D_+(z=0)=1$, and $\delta_{m,\mathrm{ad}}$ denotes the adiabatic (growing) part of the total-matter perturbations. On the large scales used in our analysis ($k\ll k_{\mathrm{fs}}$), non-growing contributions are already subdominant at $z_{\mathrm{ini}}$, so $\delta_{m,\mathrm{ad}}(k,z_{\mathrm{ini}})$ can be taken from the \texttt{CLASS} total-matter transfer function. A simple consistency check is that $T_{\delta_m}(k,z)/D_+(z)$ is time-independent across two nearby redshifts around $z_{\mathrm{ini}}$, confirming isolation of the adiabatic growing mode.

Throughout, we work to leading order in the free-streaming (pressure) expansion, retaining only terms proportional to $(k/k_{\mathrm{fs}})^2$ and neglecting higher orders $\mathcal{O}\!\big[(k/k_{\mathrm{fs}})^4\big]$. We also neglect the adiabatic decaying mode in the total-matter perturbations, as it rapidly becomes irrelevant after matter domination and has no observable impact at the redshifts considered. While we display results using the EdS growth law, any small differences between $\Lambda$CDM and EdS growth on the large, perturbative scales and redshifts of interest can be absorbed into the galaxy-bias coefficients introduced in Section~\ref{sec:linear galaxy bias}.

\begin{figure}
\centering  \includegraphics[width=1\linewidth]{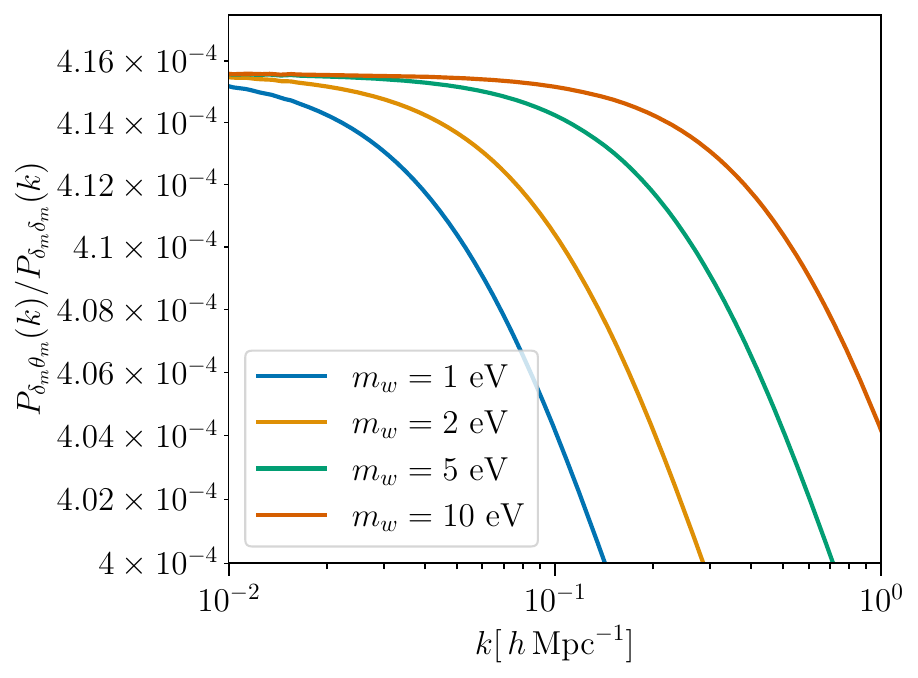}
\caption{
Scale dependence of the growth rate, defined as the ratio \(-P_{\delta\theta}(k)/P_{\delta\delta}(k)\), for MDM models with WDM particle masses \(m_w = 1, 2, 5,\) and \(10\,\mathrm{eV}\). In all cases, the warm component makes up a fixed \(15\%\) of the total dark matter density. The transfer functions are evaluated at \(z=10\) and used as the initial conditions for our fluid equations.
}
\label{fig:scale dependent growth}
\end{figure}
 
It is important to emphasize that a scale-independent growth factor $D_+(a)$, with $D_+(a) = a$ in EdS, is not valid across all scales, as the presence of the warm component leads to a suppression of growth on small scales. However, any scale-dependent behavior in the growth factor scales as $(k/k_{\mathrm{fs}})^2$, and is thus suppressed on scales larger than $k_{\mathrm{fs}}$. In Figure~\ref{fig:scale dependent growth}, we see that on perturbative scales the scale dependence of the growth rate amounts only to sub-percent level for MDM scenarios with a warm component of mass $m_w \gtrsim 10\ \mathrm{eV}$ at $z=10$, indicating that the scale-dependent growth can be safely neglected at and above this mass. We choose $z_{\rm ini} = 10$ as the initial time for evaluating the transfer functions. This choice is conservative, in the sense that evaluating them at lower redshift would further suppress any residual scale dependence and thus make our conclusions even more robust. At the same time, since galaxy formation depends on the past history of perturbations, the initial time should be set sufficiently early, when structure formation is still close to linear, which is well satisfied at $z \sim 10$.  

After obtaining the solutions for the total matter perturbations in Eq.~\eqref{eq:total matter solns}, we substitute them into the pressure term that enters the fluid equations for the relative perturbations. Although the full solution for $\delta_m$ contains multiple contributions—organized as a series in $(k^2/k_\mathrm{fs}^2)$—it is sufficient to retain only the dominant growing mode, $\delta_m \approx a\,\delta_m^{(0)}$, when evaluating its effect on the relative modes. Including the leading pressure-sourced correction, $R_p$, would only generate contributions proportional to $k^4$, i.e., $\mathcal{O}\!\left[(k/k_\mathrm{fs})^{4}\right]$, which lie beyond our truncation and are therefore neglected. In this approximation, the pressure term acts as an external source in the equations for the relative perturbations. Solving these sourced equations then yields analytic expressions for the relative perturbations:
\begin{align}\label{eq:relative solns}
\delta_{r}^{(1)}(\bm{k},a) &= R_0(\bm{k}) - \ln{(a)}\frac{R_p(\bm{k})}{H_0} + \frac{2R_-(\bm{k}) }{\sqrt{a}H_0}  \nonumber \\
\theta^{(1)}_{r}(\bm{k},a) &=  \frac{R_p(\bm{k}) }{\sqrt{a}} + \frac{R_-(\bm{k})}{a} .
\end{align}
We obtain a logarithmically growing relative-density mode which has not been considered in the literature before. One can verify the accuracy of our approximations with the help of numerical results obtained from \texttt{CLASS}. To do so, we extract transfer functions for $\theta_r$ at two different redshifts, denoted as $z_1,z_2$. Since Eq.~\eqref{eq:relative solns} includes two independent modes for $\theta_r$, namely $R_p(\bm{k})$ and $R_-(\bm{k})$, evolving respectively as $1/\sqrt{a}$ and $1/a$, this pair of snapshots allows us to independently determine the two modes. We repeat the same procedure for a second redshift pair, $z_2,z_3$. If the time evolution derived in Eq.~\eqref{eq:relative solns} is correct, the $k$-dependent mode amplitudes $R_p^{z_1z_2}$, $R_p^{z_2z_3}$ and $R_-^{z_1z_2}$, $R_-^{z_2z_3}$ extracted from each redshift pair should match each other precisely. 
Similarly, once $R_p$ and $R_-$ are known, one can infer $R_0$ from the numerical Boltzmann output for $\delta_r$, using Eq.~\eqref{eq:relative solns}, and verify consistency by comparing $R_0^{z_1 z_2}$ and $R_0^{z_2 z_3}$ obtained at different redshifts.

\begin{figure}[t]
\begin{subfigure}{.5\textwidth}
  \centering  \includegraphics[width=1\linewidth]{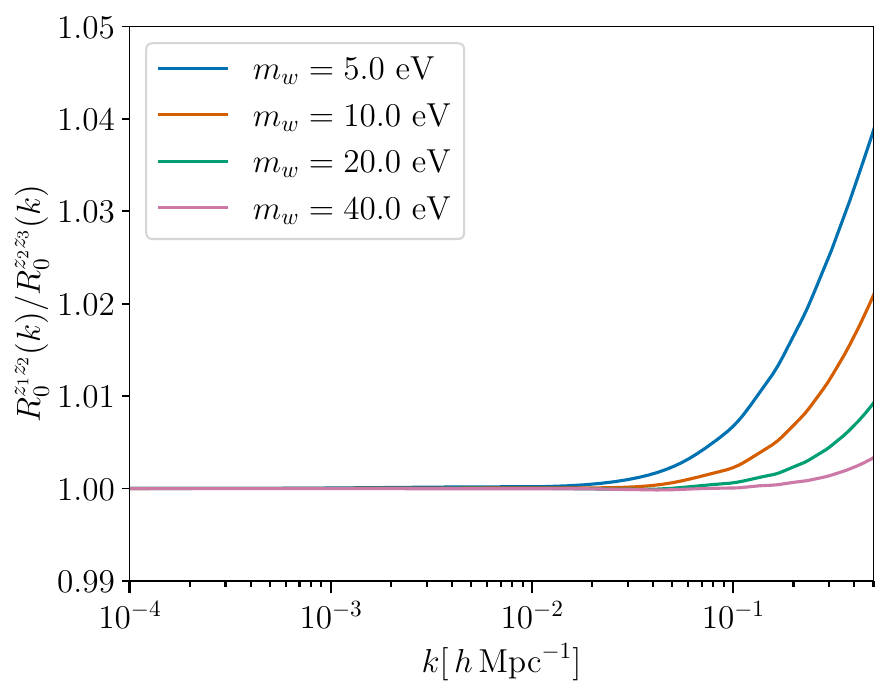}
\end{subfigure}%
\begin{subfigure}{.5\textwidth}
  \centering
  \includegraphics[width=1\linewidth]{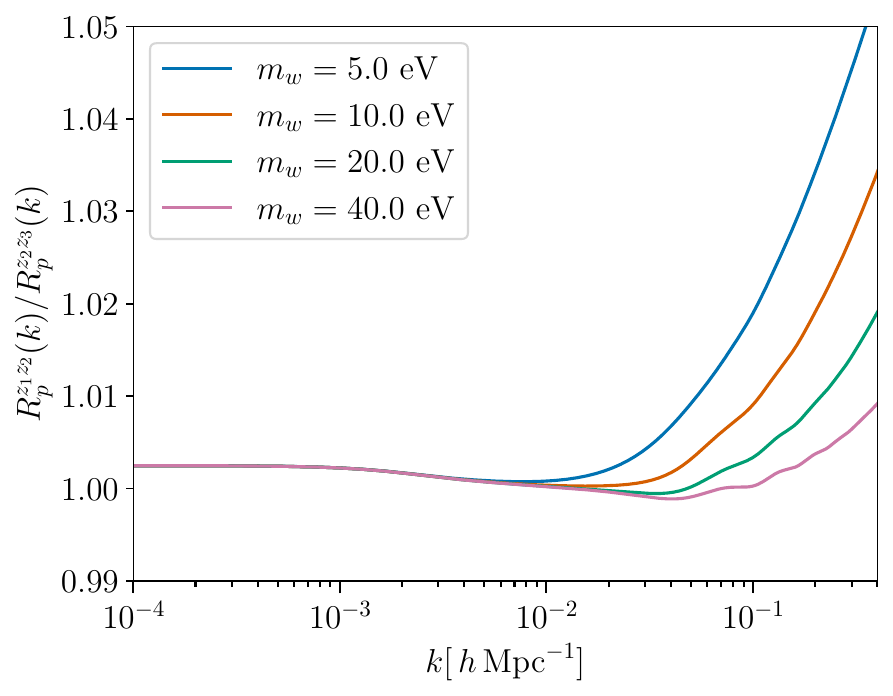}
\end{subfigure}
\caption{The left panel shows the ratio of two independently obtained $R_0$ (relative density) modes with $m_w=\, 5, \, 10, \, 20, \, 40 \ \mathrm{eV}$, which use transfer functions evaluated at $z_1,z_2,z_3=10,15,20$ respectively. The right panel shows the corresponding ratio for $R_p$ modes.}
\label{fig:time evolution relative modes}
\end{figure}

In Figure~\ref{fig:time evolution relative modes} we demonstrate this procedure, finding excellent agreement between the theoretical predictions and numerical results in an Einstein-de Sitter universe for sufficiently large scales. 
While percent-level deviations appear at small scales---more pronounced in the $R_p$ mode---we emphasize that our analysis adopts a conservative choice for $k_{\mathrm{max}}$, as discussed in Sec.~\ref{sec:Fisher}. This ensures that we remain within the regime where the analytic mode decomposition is reliable and the underlying approximations are fully justified. Although probing mixtures with smaller masses may still be feasible, we restrict our analysis to $m_w \ge 10 \ \mathrm{eV}$ as the lightest mass considered, which results in deviations of roughly $3\%$ for $R_p$ and $1\%$ for $R_0$ on perturbative scales.

Note that the growth functions in \refeq{relative solns} will change when moving from Einstein-de Sitter to $\Lambda$CDM. However, this change is numerically small, especially as we will consider observations at $z\gtrsim 1$.

\begin{figure}
    \centering
    \includegraphics[width=0.7\linewidth]{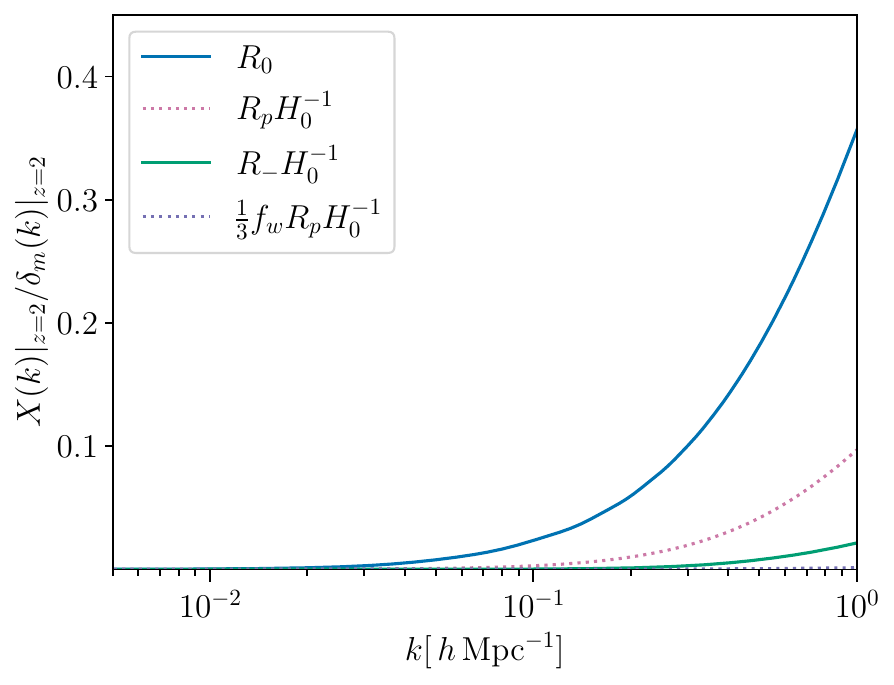}
    \caption{Different modes appearing in the two-fluid system as a function of scale $k$ at redshift $z=2$, relative to the total-matter growing mode. We consider a MDM scenario with $m_w = 10 \ \mathrm{eV}$ and $f_w=15\%$. Solid lines indicate positive values, while short-dashed lines indicate negative ones. All modes are determined from the \texttt{CLASS} output as described in the text. The subleading constant total matter mode  (purple dashed line) is the most suppressed mode among these. The detailed description of the \texttt{CLASS} settings are presented in Appendix~\ref{appendix:CLASS}.}
    \label{fig:hierarchy modes}
\end{figure}

\subsection{Free-streaming power counting}
We now have the evolution of all the relevant modes necessary for an accurate description of a two-fluid system consisting of one cold and one warm component. However, before continuing it is worth discussing the hierarchy among these modes.
To do so, we introduce a small parameter in our theory, related to the free-streaming scale $k_{\rm fs}$ of the particle of interest:
\begin{equation}
\epsilon_{\rm fs}(a,k)\equiv \frac{c_s^2(a)\,k^2}{\mathcal{H}^2(a)}=\left(\frac{k}{k_{\rm fs}(a)}\right)^2,\qquad
k_{\rm fs}(a)=\frac{\mathcal{H}(a)}{c_s(a)}\,.
\end{equation}
For reference, we can calculate, for the lowest mass we consider ($m_w=10 \; \rm eV$), the maximum value which $\epsilon_{\rm fs}$ can attain. Using Eq.~\eqref{eq:free streaming def}, which applies to sterile neutrinos which share the same relic temperature with standard model neutrinos, we have at $z=z_{\rm ini}$
\begin{align}
\epsilon_{\rm fs} (k=k_{\rm max}, \; m_w = 10 \, {\rm eV}, \; z= z_{\rm ini} ) 
= \left[ \frac{k_{\rm max} (1 + z_{\rm ini})^2 \, h^{-1}\,\mathrm{Mpc}}{6.67 \; \Omega_m^{1/2} \;(1+z_{\rm ini})^{3/2}} \right]^2\,.
\end{align}
Adopting $z_{\rm ini} = 10$ and $\Omega_m = 1$ we obtain maximum values of
\begin{equation}
\epsilon_{\rm fs} (k_{\rm max}, \;  10 \, {\rm eV}, \;  10 )\;\simeq\; 0.25\,
\left(\frac{k_\mathrm{max}}{h\,{\rm Mpc}^{-1}}\right)^2\;
\;=\;
\begin{cases}
0.007 & (k_\mathrm{max}=0.16~h\,{\rm Mpc}^{-1}) ~\ \mathrm{DESI}\\
0.02 & (k_\mathrm{max}=0.30~h\,{\rm Mpc}^{-1})~\ \mathrm{PFS}\,,
\end{cases}
\nonumber
\end{equation}
where we have adopted survey-specific values of $k_\mathrm{max}(z) = 0.3 k_\mathrm{nl}(z)$ defined in Sec.~\ref{sec:Fisher}. Clearly, for the mass range considered we are in the $\epsilon_{\rm fs} \ll 1$ regime.

Hence, we organize an expansion in $\epsilon_{\rm fs}$ for each perturbative order $n$,
\begin{equation}
\delta_m^{(n)}=\delta_{m,0}^{(n)}+\delta_{m,\epsilon}^{(n)}+\mathcal{O}(\epsilon_{\rm fs}^2)\,,\qquad
\theta_m^{(n)}=\theta_{m,0}^{(n)}+\theta_{m,\epsilon}^{(n)}+\mathcal{O}(\epsilon_{\rm fs}^2)\,.
\end{equation}
At linear order in an EdS background, the total and relative solutions read:
\begin{align}
\delta_m^{(1)}(k,a) &= D_+(a)\,\delta_m^{(0)}(k) \;-\; \frac{f_w}{3H_0}\,R_p(k) \;+\; \mathcal{O}(\epsilon_{\rm fs}^2)\,,\qquad D_+=a\,,\label{eq:lin-m}\\
\delta_r^{(1)}(k,a) &= R_0(k) \;-\; \ln a\,\frac{R_p(k)}{H_0}+ \frac{R_-(\bm{k}) }{\sqrt{a}H_0}\,,\qquad
\theta_r^{(1)}(k,a)=\frac{R_p(k)}{\sqrt{a}} 
+ \frac{R_-(\bm{k})}{a}\,.\label{eq:lin-r}
\end{align}
The pressure-sourced particular mode satisfies
\begin{equation}
R_p(k)\propto H_0^{-1}c_s^2(a)\,k^2\,\delta_m^{(0)}(k)\,,
\end{equation}
so that $R_p/(H_0\delta_m)=\mathcal{O}(\epsilon_{\rm fs})$ up to factors of order unity. The homogeneous relative mode $R_0$ is also suppressed with respect to $\delta_m$, but less strongly than $R_p$.  Over the scales of interest we find an amplitude ratio
\begin{equation}
C(a,k;m)\equiv \frac{H_0R_0}{R_p}\simeq 10\,,
\end{equation}
with almost no dependence on the warm-component mass $m_w$. However one should keep in mind that this relation between the relative modes is most likely model dependent.
In order to obtain a systematic counting,
we make the ansatz that $R_0$ is $\mathcal{O}(\epsilon_{\rm fs}^{1/2})$. This counting is accurate when $R_p/H_0R_0 = C^{-1} = \epsilon_{\rm fs}^{1/2}$, or
$\epsilon_{\rm fs} \sim 0.01$. For larger values of $\epsilon_{\rm fs}$, this counting underestimates the importance of $R_0$; however, our minimal mass value excludes cases with $\epsilon_{\rm fs}$ much larger than $0.01$. Therefore our ansatz for $R_0$ is conservative for the mass regime we work in.
Hence,
\begin{equation}
\delta_r^{(1)}=R_0+\mathcal{O}(\epsilon_{\rm fs})\,,\qquad \theta_r^{(1)},\,u_r^{(1)}=\mathcal{O}(\epsilon_{\rm fs})\,.
\end{equation}

As illustrated in Figure \ref{fig:hierarchy modes}, the most dominant mode is, as expected, the growing total matter density mode, which serves as the reference here. This is followed by the constant and the logarithmically growing relative density modes, while the $R_-$ mode present in the relative perturbations (the green solid curve), as well as the sub-leading total matter mode (the purple dashed curve) influenced by the pressure term, are much more suppressed. Consequently, we will omit the $R_-$ contribution in the following treatment due to its strongly decaying time dependence. We leave the discussion of the $R_p$ term appearing in $\delta_m$ to Sec.~\ref{sec:linear galaxy bias}.

\section{Linear Galaxy Bias}\label{sec:linear galaxy bias}
Since we cannot directly observe matter density fluctuations, but instead observe tracers such as galaxies, we need a formalism that systematically connects the galaxy density field to the underlying matter density field. On large scales, this connection can be established using the perturbative bias expansion, as reviewed in \cite{Desjacques_2018}:
\begin{align}
    \delta_g(\bm{k},\eta) = \sum_\mathcal{O} b_\mathcal{O}(\eta) \mathcal{O}(\bm{k},\eta) .
\end{align}
Here, the $\mathcal{O}$ represent scalar contractions of the second derivative of the gravitational potential, $\partial_i\partial_j\Phi$, which is the leading local gravitational observable at each perturbative order. The coefficients $b_\mathcal{O}$ are free parameters associated with each operator in the expansion and must be determined from observational data.

This framework offers several important advantages. Since the bias expansion includes all local gravitational observables (up to a given order in perturbation theory), it does not require detailed assumptions about the astrophysical processes that govern galaxy formation and evolution. Instead, the complex small-scale physics is absorbed into a finite number of bias parameters, which can be systematically organized and constrained using data or simulations. Moreover, the expansion is consistent with cosmological perturbation theory and,
on large scales where perturbation theory remains valid,
allows for analytic control over predictions for observables such as the galaxy power spectrum and bispectrum. 

Nonetheless, the approach has intrinsic limitations. The perturbative expansion breaks down on small, nonlinear scales where higher-order terms are no longer suppressed and non-perturbative effects dominate. Additionally, the bias coefficients are not directly observable and must be treated as nuisance parameters or fit empirically, which may introduce partial degeneracies and degrade cosmological parameter inference. Despite these caveats, the perturbative bias expansion remains a robust and the most widely used framework for interpreting galaxy clustering measurements in the linear and mildly nonlinear regime.

\subsection{Linear Bias Relation}

In the $\Lambda$CDM (or more generally quintessence-CDM) case, the matter field is treated as a single comoving fluid encompassing all relevant degrees of freedom. Consequently, all operators depend on a single mode: the adiabatic growing mode. The adiabatic decaying mode, which starts to decay shortly after matter domination, is generally neglected as we have done here [Eq.~\eqref{eq:total matter solns}].

In our case, however, we consider the evolution of two distinct matter components that are coupled through gravity, which is analogous to the baryon-CDM mixture considered in \cite{2009ApJ...700..705S,2010PhRvD..82h3520T,2015JCAP...05..019L,Schmidt_2016,Rampf:2020ety}. As discussed in Section~\ref{sec:linear matter}, we neglect the decaying relative mode $R_-$ and the constant total matter mode as they are subdominant.
This leaves us with two additional degrees of freedom in the bias expansion: a constant relative-density mode, $R_0$, and a relative-velocity and relative-density mode, $R_p$, which arises due to the effective pressure of the warm component. 

The linear-order expression for the galaxy density field can then be written as:\footnote{Strictly, the relative modes $R_0$ and $R_p$ should be evaluated at the initial Lagrangian position $\bm{q}$ corresponding to a time at high redshift, consistent with the validity of our fluid description deep in the matter-dominated era. However, the distinction between Lagrangian and Eulerian coordinates is higher order in perturbation theory, so we can evaluate all fields at the Eulerian position $\bm{x}$ here. The distinction will become significant when considering nonlinear evolution.}
\begin{align}\label{eq:linear bias}
    \delta_g^{(1)} (\bm{x},\eta) &= b_1(\eta)\delta_m^{(1)}(\bm{x},\eta)  + b_\delta^r(\eta)R_0(\bm{x},\eta) +  b_\theta^r(\eta) R_p(\bm{x},\eta) +\epsilon(\bm{x},\eta)\,.
\end{align}
Here, $\epsilon(\bm{x},\eta)$ accounts for stochasticity, which we model as Gaussian white noise that is scale independent on large scales, as appropriate in the linear regime. 
One can argue that one should also keep track of the $R_p$-mode appearing in Eq.~\eqref{eq:total matter solns} since it carries the same $\epsilon_{\rm fs}$ power as $b_\theta^r(\eta) R_p(\bm{x},\eta)$. In this context, including the $R_p$ contribution to $\delta_m$ would require introducing an additional independent bias parameter, reflecting its distinct time evolution compared to the adiabatic growing mode. Nevertheless, this contribution is degenerate with the term proportional to the $R_p$-mode in the relative velocity divergence bias, as they share the same scale dependence. Therefore, including the $R_p$ mode in $\delta_m$ would not lead to an independent new term in the galaxy power spectrum, but would instead be absorbed into existing bias parameters. For this reason, we neglect this contribution.
See Ref.~\cite{Verdiani:2025jcf} for a bias expansion written in a different basis but equivalent at linear order, applied to a MDM scenario with dominant CDM and ALPs as a subdominant second component.

\begin{figure}
\begin{subfigure}{.5\textwidth}
  \centering  \includegraphics[width=1\linewidth]{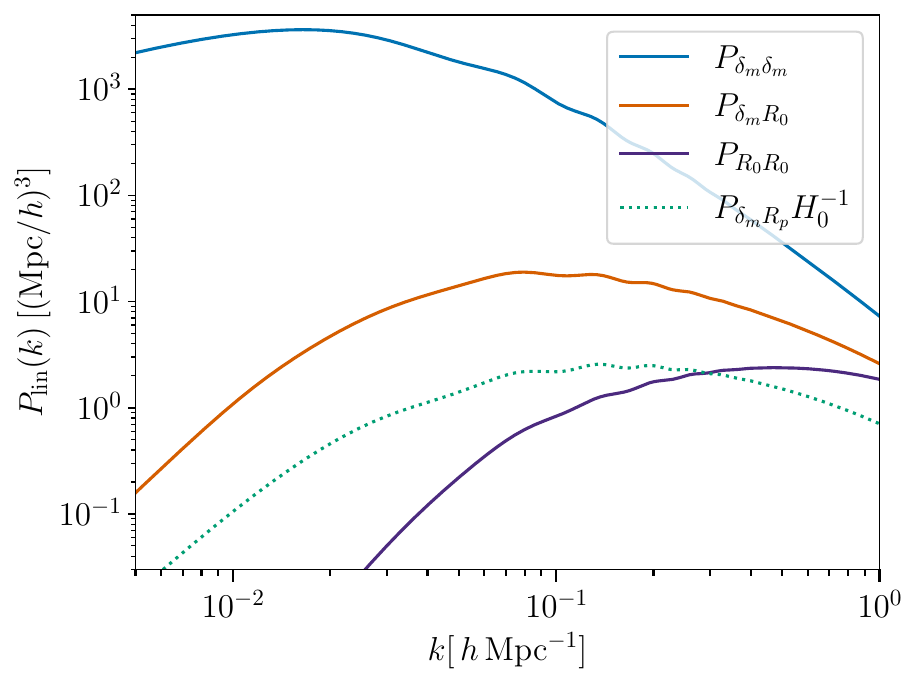}
  \label{fig:sfig1}
\end{subfigure}%
\begin{subfigure}{.5\textwidth}
  \centering
  \includegraphics[width=1\linewidth]{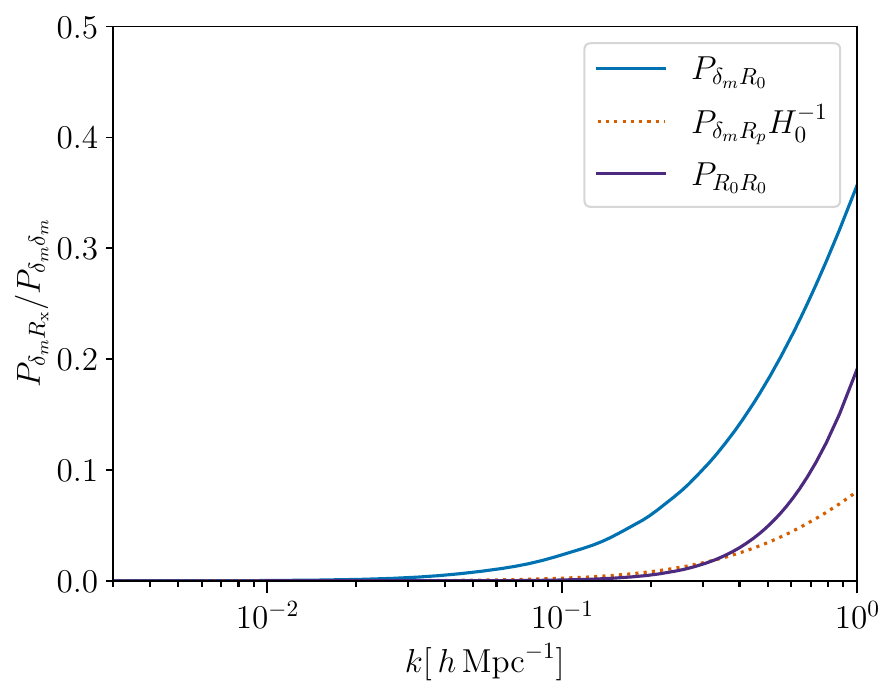}
  \label{fig:sfig2}
\end{subfigure}
\caption{The left panel shows the individual auto and cross power spectra of the total and relative matter modes at $z=2$  which appear in the galaxy power spectrum. These modes are obtained for $m_w =10\: \mathrm{eV}$ and $f_w=15 \,\%$. The right panel indicates the hierarchy between the power spectra shown in the left panel.}
\label{fig:linear ps comparison}
\end{figure}

Given the linear-order expression for the galaxy density field including the necessary relative modes, writing down the linear galaxy power spectrum is straightforward:
\begin{align}\label{eq:galaxy PS real space}
    P_{gg}^{\mathrm{lin}}(k) = b_1^2P_{\delta_m\delta_m}^{\mathrm{lin}}(k) + 2b_1b_\delta^r P_{R_0\delta_m}^{\mathrm{lin}}(k) + 2b_1b_\theta^r P_{R_p\delta_m}^{\mathrm{lin}}(k) + \left(b_\delta^r\right)^2P_{R_0R_0}(k) + P_\epsilon\,.
\end{align}
where we have neglected the auto-correlation of the $R_p$ mode since it contributes at $\mathcal{O}(\epsilon_{\rm fs}^2)$ (although it is easy to include, and does not introduce any additional free parameters).
Figure~\ref{fig:linear ps comparison} illustrates the different contributions quantitatively. The power of $P_{\delta_mR_0}$ is around $15\%$ of $P_{\delta_m\delta_m}$ at $k \approx 0.3 \iMpch$, which is non-negligible (this statement depends on the value of the bias coefficient $b_\delta^r$, which we discuss below). The stochastic field is not correlated with the other fields, therefore it only contributes via the stochastic power spectrum $P_\epsilon \sim n_g^{-1}$ which is roughly controlled by the number density of the galaxies in a survey. 

In order to estimate the magnitude of the bias parameters, it is useful to recall their physical interpretation. Specifically, one can view the galaxy number density as a local functional of the CDM and WDM perturbations, such that $n_g(\bm{x},z) = F_g[\delta_m, \delta_r]$. Each bias parameter can then be understood as the response of $n_g$ to a variation in the corresponding operator. The bias coefficient $b_\delta^r$ defined by
\begin{align}\label{eq:bias estimate b_r}
    b_\delta^r = \frac{1}{F_g[0]} \left. \frac{\partial F_g}{\partial \delta_r} \right|_{0},
\end{align}
quantifies the response of a galaxy population to a long-wavelength perturbation in the relative density field $\delta_r$ between two species, such as cold and warm dark matter. Physically, this measures how the abundance of galaxies changes when embedded in a large-scale environment where the relative density between the components is slightly enhanced or suppressed. The response is normalized by the mean abundance $F_g[0]$, yielding a dimensionless bias parameter that characterizes the sensitivity of galaxy formation to relative fluctuations.

In practice, this bias can be estimated using the separate universe approach. In this method, a small, constant $\delta_r$ perturbation is introduced into the background density in the initial conditions of a cosmological simulation, while keeping the total matter density fixed. Simulations with slightly positive and negative values of $\delta_r$ are run, and the resulting change in galaxy abundance is measured. The derivative in Eq.~\eqref{eq:bias estimate b_r} is then extracted by comparing these outcomes, providing a direct numerical estimate of the bias of the simulated galaxy sample. This method was applied to measure the bias with respect to the baryon-CDM relative density perturbation in \cite{Barreira:2023uvp}.

While it would be interesting to likewise estimate the bias coefficients $b_\delta^r, b_\theta^r$ from full numerical simulations of MDM, here we restrict ourselves to providing analytical estimates. 
For this, it is more convenient to change our basis from $\delta_m,\delta_r$ to $\delta_c,\delta_w$.
Eq.~\eqref{eq:linear bias} can be equivalently written as:
\begin{align}\label{eq:bias estimate}
    \delta_g^{(1)}(\bm{x},\eta) &= b_c \delta_c^{(1)}(\bm{x},\eta) + b_w \delta_w^{(1)}(\bm{x},\eta) \nonumber \\
    \delta_g^{(1)}(\bm{x},\eta) &= (b_c + b_w)\delta_m + (f_w b_c - b_w f_c)\delta_r.
 \end{align}
It is not trivial to estimate $b_\delta^r$ by Eq.~\eqref{eq:bias estimate} ; however, we can look at a limiting scenario for the warm component to obtain a rough estimate of the allowed values of $b_\delta^r$. First, if $m_w \approx 0$, the warm component will free-stream on all scales, and will not participate in the formation of nonlinear structure. This implies $b_w \approx 0$ and $b_c \approx b_1$, leading to the approximation 
\begin{align}\label{eq:br limits}
b_\delta^r \approx f_w b_1
\end{align} We stress, however, that this linear scaling is only expected to hold in the small-$f_w$ regime where the warm component represents a subdominant fraction and its impact on the cold component's clustering can be treated perturbatively.
For higher warm masses the allowed $f_w$ is less well constrained, and the relation $b_\delta^r \propto f_w$ need not apply (e.g.\ if the warm component clusters non-negligibly or backreacts on $b_c$). In practice, obtaining a precise estimate of $b_\delta^r$ in the large-$f_w$ regime requires dedicated numerical simulations and a direct measurement of the relevant bias coefficients which we leave to a future work.

After setting the physical limits on $b_\delta^r$, we can relate it to $b_\theta^r$ using Eq.~\eqref{eq:relative solns}.  
A nonzero $\theta_r(\mathbf{x}, z)$ is associated with a nonzero relative density perturbation $\delta_r$ derived above, which satisfies
\begin{equation}
\theta_r \approx  \, H_0^{-1} \, \delta_r,
\end{equation}
We can then estimate
\begin{equation}\label{eq:relation br bt}
b_\theta^r \equiv \frac{1}{\bar{n}_g} \frac{\partial \bar{n}_g}{\partial \theta_r(z)} 
\sim b_\delta^r \, \frac{\partial \delta_r}{\partial \theta_r(z)} 
\approx  \, H_0^{-1} \, b_\delta^r .
\end{equation}
Notice that $b_\theta^r$ has dimensions of length, which is taken care of by the factor of $H_0^{-1}$.
Assuming $b_\delta^r$ is taking the value allowed by our approximation, we therefore expect $b_\theta^r$ to be of order $H_0^{-1}$ as:
\begin{align}\label{eq:bt limits}
b_\theta^r \approx H_0^{-1}f_wb_\delta^r\,.
\end{align}

\subsection{Linear Velocity Bias}

To accurately model the distribution of galaxies observed in redshift space, it is essential to account for galaxy velocities, as they affect the observed power spectrum through redshift-space distortions (RSD). In particular, when working within a two-fluid framework consisting of CDM and WDM, we must determine how perturbations in each fluid contribute to the total galaxy velocity.

In the single-fluid approximation, the equivalence principle ensures that the velocity bias between galaxies and matter vanishes at linear order. However, this no longer holds in the presence of two distinct dark matter components, where galaxies may not follow the total matter velocity exactly. This leads to a non-zero velocity bias at linear order due to the initial relative velocity between CDM and WDM. To incorporate this effect, we define the velocity bias $\bm{v}_{\rm bias}$ as the difference between the galaxy velocity $\bm{v}_g$ and the total matter velocity $\bm{v}_m$:
\begin{align}
    \bm{v}_\mathrm{bias} (\bm{x},\eta) = \bm{v}_g (\bm{x},\eta) - \bm{v}_m (\bm{x},\eta)\,.
\end{align}
Given the above discussion, we need to write the velocity bias at linear order as
\begin{align}\label{eq:velocity bias}
    \bm{v}_\mathrm{bias} (\bm{x},\eta) = \beta_r\bm{v}_r (\bm{x},\eta) + \mathrm{higher \ derivative \ terms}
\end{align}
where the higher derivative terms correspond to the velocity bias in the usual single-fluid case, which are expected to be controlled by the physical length scale $R_*$ for galaxy formation \cite{Desjacques_2018}. One usually assumes that $R_*$ is of order the nonlinear scale or smaller, so that higher-derivative terms are compareable to nonlinear corrections, and we do not consider them further here. The coefficient of the first term, $\beta_r$, can be bracketed to take values in between $(-1,1)$, through the following argument. If galaxies purely follow the CDM component, then we have $\beta_r = f_w$, while in the opposite case, where galaxies only follow the warm component, we have $\beta_r = -f_c$. Given that the fractions can lie between 0 and 1, we obtain the above quoted overall range for $\beta_r$. This range could be narrowed more in a concrete analysis by factoring in the allowed region of $f_w$.

Now including the first term in Eq.~\eqref{eq:velocity bias}, we obtain the linear galaxy power spectrum in redshift space as follows:
\begin{align}\label{eq:PS redshift}
P_{gg}^{s,\mathrm{lin}}(k,\mu) = \left(b_1 + f \mu^2 \right) 
& \left[ 
(b_1 + f \mu^2) P_{\delta_m \delta_m}(k) 
+ 2 b_\delta^r P_{\delta_m R_0}(k) 
+ 2b_\theta^r  P_{\delta_m R_p}(k) \right. \nonumber \\
& \left. - \frac{\beta_r}{aH} \mu^2 P_{\delta_m R_p}(k) 
\right] +{b_\delta^r}^2P_{R_0R_0(k)}  + P_\epsilon,
\end{align}
where the first term reproduces the well known Kaiser formula \cite{Kaiser}, $f \equiv d\ln D_+/d\ln a$ being the linear growth rate  (not to be confused with $f_w$) and $\mu$ the cosine of the angle between line of sight direction and wavevector $\bm{k}$. The last term is the contribution due to the linear velocity bias we have discussed above. \refeq{galaxy PS real space} and \refeq{PS redshift} give our complete model describing the linear statistics of galaxy clustering both in real and redshift space.

\section{Fisher Forecast}\label{sec:Fisher}
In this section we forecast the constraining power of ongoing galaxy redshift surveys on the MDM scenario considered in this work, parametrized by the mass $m_w$ and mass fraction $f_w$ of the WDM component. Fisher forecasting provides an efficient way to estimate expected parameter constraints for a given survey setup. A particular focus is on the effect of the relative density ($R_0$) and velocity ($R_p$) modes in the two-fluid cosmology on the constraints, obtained by marginalizing over the corresponding bias coefficients $b_\delta^r,\  b_\theta^r$. We first consider the real-space galaxy power spectrum, before moving on to the redshift-space case.

We focus on the Subaru Prime Focus Spectrograph (PFS)~\cite{PFS} and DESI~\cite{DESI:2016fyo} surveys. PFS is smaller volume but probes higher redshifts than DESI. This comparison provides insights into how the impact of relative modes depends on survey redshift coverage and observational characteristics, and allows us to identify which survey is most sensitive to different mass ranges of warm particles.

\subsection{Real Space Analysis}

In a Fisher analysis, we estimate how well a set of model parameters can be constrained by examining the expected curvature of the likelihood function around its maximum. Under the assumption that the likelihood is Gaussian and that the data covariance matrix is diagonal and does not depend on the parameters being constrained, the Fisher information matrix is given by:
\begin{align}\label{eq:fisher formula}
    F_{\alpha \beta} = \sum_{k_i} \frac{\partial P_g(k_i)}{\partial \theta_\alpha} \frac{1}{\mathrm{Var}[P_g(k_i)]} \frac{\partial P_g(k_i)}{\partial \theta_\beta},
\end{align}
where $\{\theta_\alpha\}$ is the set of model parameters, $P_g(k_i)$ is the predicted galaxy power spectrum for a wavenumber bin $i$ centered at $k_i$, and $\mathrm{Var}[P_g(k_i)]$ is the variance of the measured power spectrum in a given $k$-bin.

This expression relies on several key assumptions. First, we assume that the likelihood of the observables is Gaussian with diagonal covariance, which is a good approximation on linear scales where fluctuations are nearly Gaussian. To remain in this regime, on which our linear model for $P_g(k)$ also relies, we restrict our analysis to wavenumbers $k \lesssim 0.3\,k_{\mathrm{nl}}(z)$, where nonlinear corrections are still small and perturbation theory remains reliable. Here, $k_{\mathrm{nl}}(z)$ denotes the nonlinear scale at redshift $z$, defined implicitly by the condition
\begin{align}
\Delta^2(k_{\mathrm{nl}}, z) \equiv \frac{k_{\mathrm{nl}}^3 P_m(k_{\mathrm{nl}}, z)}{2\pi^2} = 1,
\end{align}
where $P_m(k, z)$ is the linear matter power spectrum and $\Delta^2(k, z)$ is the variance of (linear) density perturbations.  This criterion defines $k_\mathrm{max}$ in our analysis of the two surveys. 

We focus on the redshift bin around $z=2$ for PFS and $z=1$ for DESI, correspondingly adopting $k_\mathrm{max}=0.30\,h\,\mathrm{Mpc}^{-1}$ for PFS and $k_\mathrm{max}=0.16\,h\,\mathrm{Mpc}^{-1}$ for DESI, given that $k_{\mathrm{nl}}(z=2) = 1\,h\,\mathrm{Mpc}^{-1}$ and $k_{\mathrm{nl}}(z=1) = 0.55\,h\,\mathrm{Mpc}^{-1}$. While these $k_\mathrm{max}$ values may appear somewhat optimistic, it is worth noting that contributions from the relative modes to the linear galaxy power spectrum scale as $(k/k_\mathrm{fs})^2$, which is already at the order of one-loop corrections. 
These scales can be safely modeled using (EFTofLSS) at one-loop order; however, going to one-loop order would introduce many additional free parameters in the galaxy power spectrum, which would ultimately degrade the constraining power of our analysis.

In addition to assuming a Gaussian likelihood for the observables, the Fisher matrix also approximates the posterior distribution of the parameters as Gaussian. This is a less accurate assumption, as posteriors can exhibit non-Gaussian features such as skewness or curved degeneracies, especially when parameters are not tightly constrained. As a result, this approximation can lead to underestimated uncertainties and should be interpreted with the appropriate caution.

Finally, the forecast depends on survey-specific inputs such as volume, redshift range, and galaxy number density. To see this dependence let us have a closer look at the variance \cite{Feldman_1994}:
\begin{align}
    \mathrm{Var}[P_g(k_i)] = \frac{1}{N_{k_i}}\left[P_g(k_i)\right]^2
\end{align}
where $N_{k_i}$ is the number of Fourier modes within each bin for which we evaluate the power spectrum in Eq.\eqref{eq:galaxy PS real space}. Here we are not bothered with the details of the survey geometry and assume $N_{k_i}$ is the same as in case of a cubic survey with with a volume $V_{\mathrm{Survey}}$. In this simplified case one directly obtains
\begin{align}\label{eq:number of modes}
    N_{k_i} = \frac{1}{2}\frac{4\pi k_i^2 \Delta k}{k_F^3} = \frac{V_\mathrm{Survey}}{4\pi^2}k_i^2 \Delta k
\end{align}
where, $k_F = 2\pi / V_\mathrm{survey}^{1/3}$ denotes the fundamental wavenumber of the survey volume. The factor of $1/2$ accounts for the reality of the density field, which implies that only half of the Fourier modes within a given shell are independent. For our analysis, we adopt $V_\mathrm{survey} = 5.3\,(\mathrm{Gpc}/h)^3$ and a mean galaxy number density of $\bar{n}_g = 3 \times 10^{-4}\,(\mathrm{Mpc}/h)^3$ by combining the two largest redshift bins of the PFS survey~\cite{PFS}. Similarly, for DESI, we focus on four redshift bins centered around $z_{\mathrm{survey}} = 1.1$. In this case, we take $V_\mathrm{survey} = 40\,(\mathrm{Gpc}/h)^3$, $\bar{n}_g = 3 \times 10^{-4}\,(\mathrm{Mpc}/h)^3$.

In our analysis, we consider the following parameter vector:
\begin{align}\label{eq:parameters}
\vec{\theta} = \{b_1, b_\delta^r,b_\theta^r, \beta_r, f_w, A_s, n_s, H_0, \Omega_{\mathrm{DM}}h^2, \Omega_bh^2\}, 
\end{align}
where the parameters—--$A_s$, $n_s$, $H_0$, $\Omega_{\mathrm{DM}}$, and $\Omega_b$—--are standard cosmological parameters describing the primordial power spectrum, background expansion and total dark matter/baryon energy densities respectively. $b_1$, $b_\delta^r, b_\theta^r$ and $\beta_r$ are the bias coefficients which we treat as free parameters, and $f_w$ is the fraction of the warm component which we will constrain for individual WDM masses. 

To improve parameter constraints and break degeneracies, we combine the Fisher matrix from the galaxy survey with a prior Fisher matrix from the Planck CMB experiment. The total Fisher matrix is  
\begin{align}
F^{\mathrm{Full}}_{\alpha\beta} = F_{\alpha\beta} + F^{\mathrm{CMB}}_{\alpha\beta},
\end{align}
where $F^{\mathrm{CMB}}_{\alpha\beta}$ is derived from the analysis of Planck data~\cite{Tristram_2024, 2020}. This combination incorporates the additional constraining power of the early universe on the standard $\Lambda$CDM parameters. 

In our analysis, the CMB information is used solely to constrain the $\Lambda$CDM sector, not the MDM parameters. This is justified because Planck probes scales much larger than $k_{\mathrm{fs}}$ at recombination, so the CMB effectively measures a single CDM component. While, in principle, CMB data could provide some sensitivity to $f_w$ through changes in $N_\mathrm{eff}$, this effect is negligible in our case: the sterile neutrinos are produced non-thermally and are already non-relativistic by recombination, making their contribution to $N_\mathrm{eff}$ minimal. 

While the derivatives with respect to bias parameters in \refeq{fisher formula} can be performed analytically, in general these derivatives need to be evaluated numerically. In particular, the derivative with respect to $f_w$ must be computed numerically, since $f_w$ also modifies the transfer functions of the individual fluid components in a nontrivial way. This arises due to the nontrivial back-reaction between the CDM and the additional fluid component, which modifies the evolution of perturbations in both sectors.\footnote{ A recent study~\cite{lee2025rapidaccuratenumericalevolution} introduced a modified Boltzmann solver called \texttt{CLASSIER}, an extension of the \texttt{CLASS} code we also use. \texttt{CLASSIER} is designed to handle non-cold relics with improved accuracy and may be a useful tool for cross-checking our results, especially in models with nontrivial multi-component dynamics. }

\begin{figure}
    \centering
    \includegraphics[width=1\linewidth]{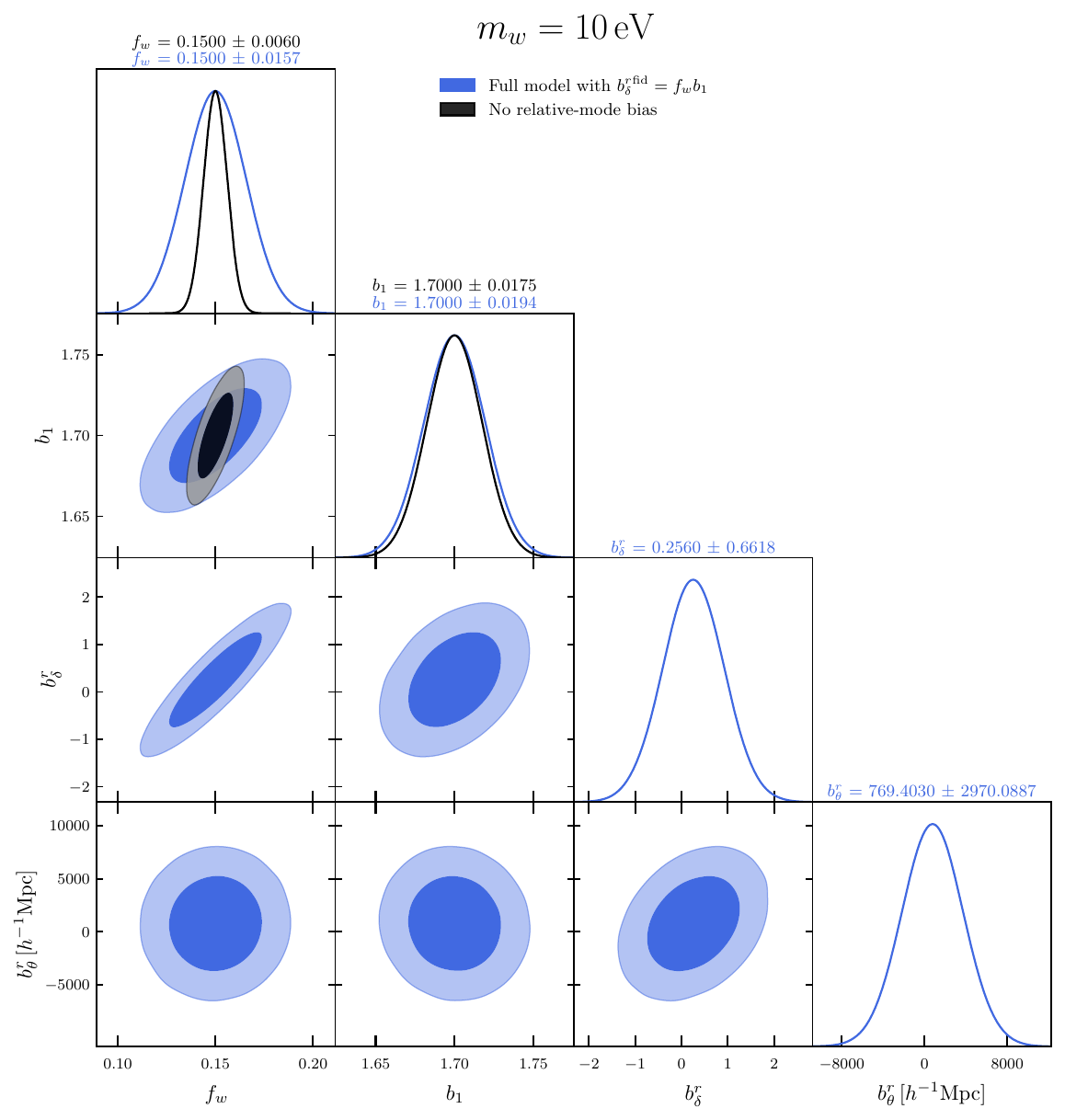}
    \caption{Fisher parameter posteriors for $b_1$, $f_w$, $b_\delta^r$ and $b_\theta^r$, in an MDM scenario with $m_w= 10\:\rm eV$ and $f_w^{\mathrm{fid}} = 15\, \%$, marginalized over all other parameters listed in Eq.~\eqref{eq:parameters}. We assume specifications for the PFS survey, set $k_{\mathrm{max}}=0.3\iMpch$, and adopt the linear galaxy power spectrum at $z=2$ from Eq.~\eqref{eq:galaxy PS real space} for two different fiducial parameter cases:
        including relative density and velocity perturbation with bias parameters corresponding to approximations in Eq.~\eqref{eq:br limits} and  Eq.~\eqref{eq:bt limits} respectively (blue), and with relative-mode bias parameters set to zero (black).}
    \label{fig:corner 10ev matteronly}
\end{figure}

In Fig.~\ref{fig:corner 10ev matteronly}, we present the constraining power of the linear galaxy power spectrum for the PFS survey, comparing two models. The first (black contours) corresponds to a simplified scenario including only the total matter power spectrum contribution, i.e. it assumes that the bias coefficients for the relative modes vanish: $b_\delta^r = 0 = b_\theta^r$. The second (blue) incorporates the full model described in Eq.~\eqref{eq:galaxy PS real space}. For clarity, we omit the cosmological parameters listed in Eq.~\eqref{eq:parameters} from the plot; however, the full posterior distribution, including all parameter degeneracies, is provided in Appendix~\ref{appendix:Full results}. Fiducial cosmological parameters are taken from~\cite{Tristram_2024}, with $b_1^{\mathrm{fid}}$ adopted from~\cite{PFS}, and the warm dark matter fraction set to $f_w^{\mathrm{fid}} = 0.15$ following~\cite{Peters:2023asu}. 
We impose a Gaussian prior on $b_\delta^r$ with a standard deviation $\sigma_{b_\delta^r} = b_1^{\mathrm{fid}}$, motivated by the expectation that $b_\delta^r$ should not exceed the total bias by a large factor. Similarly, since $b_\theta^r$ is related to $b_\delta^r$ by a factor of $H_0^{-1}$, we apply a Gaussian prior with $\sigma_{b_\theta^r} = H_0^{-1} b_1^{\mathrm{fid}}$.

As seen in the figure, both models (no relative-mode bias vs. the full model) exhibit a positive correlation between $f_w$ and $b_1$; however, the degeneracy is more pronounced when relative-mode bias is allowed. This reflects a greater degree of compensation between the two parameters in the full model. This pattern can be understood from the parameter-response of the galaxy power spectrum: around the fiducial point we find $\partial P_{gg}/\partial f_w < 0$, while $\partial P_{gg}/\partial b_\delta^r > 0$, so increasing $f_w$ can be (partly) compensated by increasing $b_\delta^r$, yielding a positive correlation between these parameters. Moreover, $\partial P_{gg}/\partial b_\theta^r < 0$, implying a positive correlation between $b_\theta^r$ and $b_\delta^r$ and a negative correlation between $b_\theta^r$ and $f_w$, consistent with the trends visible in Fig.~\ref{fig:corner 10ev matteronly}. The constraints on the warm component are generally strong for a 10~eV sterile neutrino in both cases, but introducing the additional nuisance parameters $b_\delta^r$ and $b_\theta^r$ leads to a significant weakening of the constraints on $f_w$—by roughly a factor of 2.5—compared to the simpler model with only the adiabatic growing mode. This degradation is primarily driven by a significant degeneracy between $f_w$ and $b_\delta^r$. The strong correlation between the parameters also suggests that neglecting the relative-mode bias in our model would induce a shift in the inferred value of $f_w$; a detailed calculation of this systematic shift will be presented explicitly in the next subsection~\ref{sec:RSD fisher}.
Given the amplitude hierarchy between $P_{\delta_m \delta_m}$ and $P_{\delta_m R_0}$, including additional observables may help break this degeneracy and better exploit the signal from relative perturbations—particularly for lighter secondary components.

We also observe that the constraint on $b_\theta^r$ is fully prior-dominated, while $b_\delta^r$ remains constrained by the data in the 10~eV case. Different fiducial choices for $b_\delta^r$ slightly affect the posterior for $f_w$, though this sensitivity is modest compared to the dominant effect of including the relative modes in the bias expansion.
Lastly, we note that the constraints on $f_w$ are not sensitive to the choice of $f_w^{\mathrm{fid}}$, but they do depend on $b_1^{\mathrm{fid}}$, which significantly impacts the signal-to-noise ratio.

\begin{figure}
    \centering
    \includegraphics[width=1\linewidth]{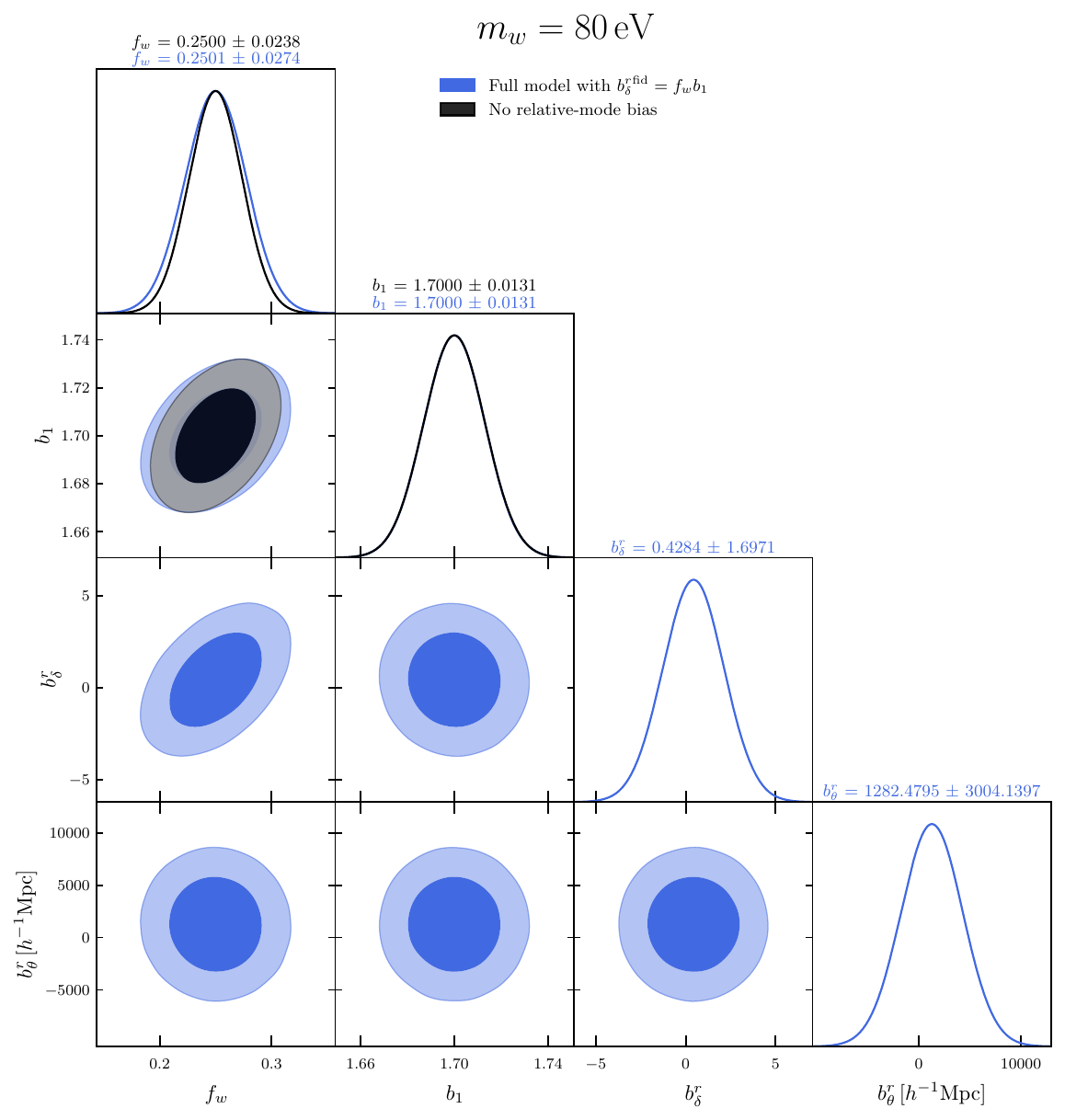}
\caption{Same as Fig.~\ref{fig:corner 10ev matteronly}, but for a WDM mass of $80\:\mathrm{eV}$ instead of $10\:\mathrm{eV}$.
We adopt the same priors and fiducial values as in Fig.~\ref{fig:corner 10ev matteronly}, except for $f_w^{\mathrm{fid}} = 25 \, \%$  at $m_w = 80\,\mathrm{eV}$.
}
\label{fig:corner 80ev matter vs full}
\end{figure}

On the other hand, as the mass of the warm component is raised, the relative perturbations between the cold and warm particles become sufficiently small so that the data can no longer effectively constrain the nuisance parameter $b_\delta^r$ either. In Fig.~\ref{fig:corner 80ev matter vs full}, which shows the same as Fig.~\ref{fig:corner 10ev matteronly} but for $m_w = 80\:\rm eV$, we observe that the uncertainty on $b_\delta^r$ is entirely dominated by the prior, similar to $b_\theta^r$. Moreover, for WDM with this mass, the inclusion of relative modes in the model becomes significantly less impactful compared to the case of lighter warm components. The change in $\sigma_{f_w}$ is at the level of $13\%$ for 80~eV particles, and this difference will diminish further for higher masses.

\subsection{Redshift Space Analysis}\label{sec:RSD fisher}

We now extend our Fisher analysis to redshift space, where peculiar velocities of galaxies introduce anisotropies in the observed clustering pattern. This redshift space distortion (RSD) effect modifies the galaxy power spectrum by introducing a dependence not only on the wavenumber \( k \), but also on the cosine of the angle between the wavevector and the line of sight, \( \mu \equiv \cos\theta \). Consequently, the Fisher matrix formalism must be updated to reflect this anisotropic nature of the signal.

In redshift space, the Fisher matrix becomes:
\begin{align}
F_{\alpha \beta} = \sum_{k_i} \sum_{\mu_j}\frac{\partial P_g(k_i, \mu_j)}{\partial \theta_\alpha} \frac{1}{\mathrm{Var}[P_g(k_i, \mu_j)]} \frac{\partial P_g(k_i, \mu_j)}{\partial \theta_\beta},
\end{align}
where $P_g(k_i, \mu_j)$ is the redshift-space galaxy power spectrum and $\mathrm{Var}[P_g(k_i, \mu_j)]$  is the variance in a given $(k, \mu)$-bin, which takes the form:
\begin{align}
\mathrm{Var}[P_g(k_i, \mu_j)] = \frac{1}{N_{k_i\mu_j}} P_g(k_i, \mu_j)^2.
\end{align}
$N_{k_i\mu_j}$ is the number of independent Fourier modes in the bin defined by $(k, \mu)$, given by:
\begin{align}
N_{k_i\mu_j} = \frac{V_\mathrm{Survey}}{8\pi^2}k_i^2 \Delta k \Delta \mu
\end{align}
Here, we have considered joint binning in $k$ and $\mu$ (also known as ``wedges''), while one could equivalently choose Legendre multipoles $P_{g,\ell}(k)$. The advantage of the chosen binning is that the covariance remains diagonal.
We adopt 10 uniform bins in $\mu \in [0,1]$, which we find sufficient to capture the relevant angular structure; increasing the number of bins further yields negligible gain in constraining power. 

In the redshift-space analysis, we introduce the velocity bias parameter $\beta_r$ as free parameter [Eq.~\eqref{eq:velocity bias}], which accounts for a possible linear velocity bias between galaxies and the underlying dark matter distribution.
We choose a fiducial value for $\beta_r$ at the midpoint of the physically allowed range $[-f_c, f_w]$, and impose a Gaussian prior such that the $1\sigma$ uncertainty spans this interval.

For the linear growth rate, we adopt the commonly used approximation $f \approx \Omega_m^{0.55}$, which is valid within a $\Lambda$CDM cosmology. Given that we do not expect scale-dependent growth in the linear perturbative regime of interest, this approximation remains valid for MDM scenarios considered here. 
While $f$ formally depends on the total matter density $\Omega_m = \Omega_{\rm{DM}} + \Omega_b$, we do not treat $\Omega_m$ as a free parameter in our analysis. Instead, we vary $\Omega_{\rm{DM}}$ and $\Omega_b$ separately. Since $\Omega_b$ is tightly constrained by CMB observations and contributes subdominantly to $\Omega_m$, we allow $f$ to vary only with $\Omega_{\rm DM}$ in our Fisher forecasts.

\begin{figure}
  \centering  \includegraphics[width=1\linewidth]{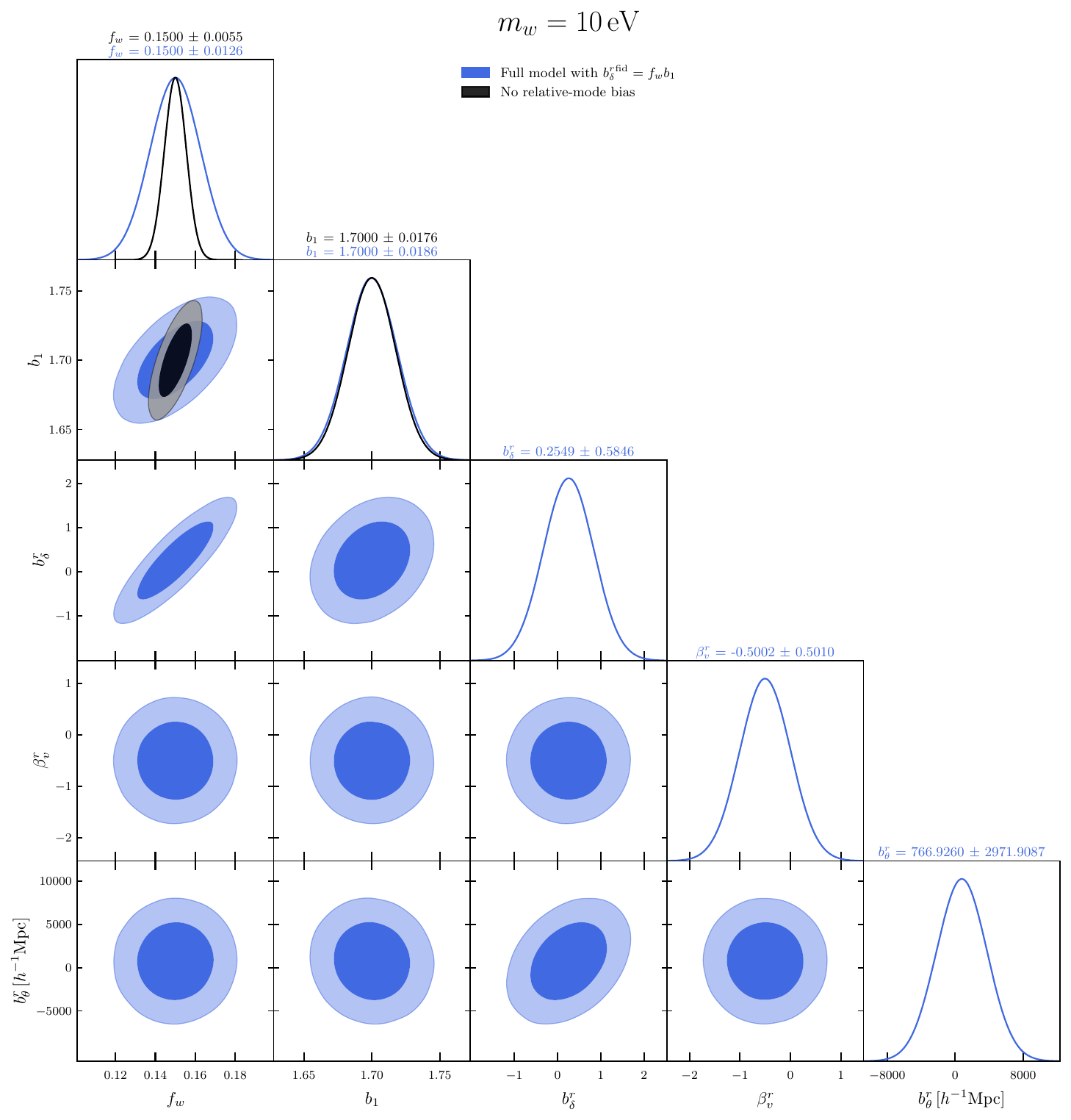}

\caption{Same as Fig.~\ref{fig:corner 10ev matteronly}, but for the redshift-space rather than real-space galaxy power spectrum. We include an additional parameter $\beta_r$ which accounts for the linear velocity bias. The fiducial value of $\beta_r$ is set to $-0.5$, with a Gaussian prior of width $\sigma_{\beta_r} = 0.5$.}
\label{fig:corner 10ev rsd}
\end{figure}

The main result of the analysis can be seen in Fig.~\ref{fig:corner 10ev rsd}. Focusing first on the constraints on $f_w$ from $P_g(k, \mu)$, we observe a slight improvement over the real-space case without RSD in both the full model---which includes relative modes and velocity bias---and the simplified model with only adiabatic modes. The trend of loosening the constraints on $f_w$ by including relative modes is also present in redshift space but the impact is not as strong as in the real space case.

The parameter $\beta_r$ remains poorly constrained due to its weak observational imprint. As shown in the figure, its posterior is largely dominated by the prior, indicating limited sensitivity in the data. Moreover, there are no significant degeneracies between $\beta_r$ and the other model parameters. As a result, treating $\beta_r$ as a free parameter does not degrade the constraints on the remaining parameters, despite its own large uncertainty. The information on $b_\theta^r$ is also prior dominated as in the real-space case. 
Another notable result is the improved constraint on $\Omega_{\mathrm{DM}}$, which arises thanks to the linear growth rate constraint from the large-scale anisotropic galaxy power spectrum.

Given the significant impact of the relative-mode bias on the forecasted error bar of $f_w$, it is worth asking what systematic bias in $f_w$ would result in a redshift-space power spectrum analysis when neglecting the relative-mode-bias contributions. To estimate this, we use the ``Fisher bias'' calculation, which yields a parameter shift of \cite{Dodelson_Schmidt_2021}
\begin{align}\label{eq:fisher bias}
\Delta f_w &= \sum_{\beta} \left(F^{-1}\right)_{f_w \beta} \, B_{\beta} ,
\end{align}
where $\left(F^{-1}\right)_{f_w \beta}$ is the $(f_w, \beta)$ element of the inverse Fisher matrix, and $B_{\beta}$ is the bias vector entry for parameter $\beta$:
\begin{align}
B_{\beta} &= \sum_{k,\mu} \frac{\partial P_g(k,\mu)}{\partial \theta_{\beta}}
\frac{1}
{\mathrm{Var}\!\left[P_g(k,\mu)\right]}
\left[P_g^{\mathrm{Full}}(k,\mu) - P_g^{\mathrm{No\ rel.-mode}}(k,\mu)
\right],
\end{align}
which is proportional to the difference between the full model and the no-relative-mode-bias case in each bin. Once $B_{\beta}$ is computed for all parameters in the no relative-bias model, the shift $\Delta f_w$ is obtained from Eq.~\eqref{eq:fisher bias} using the inverse Fisher matrix. 

For the minimum-mass case, we find that the parameter shift in $f_w$ corresponds to approximately $0.7\,\sigma_{f_w}$ for DESI. However, the amount of shift highly depends on the fiducial value of the $b_\delta^r$ and one should keep in mind that the values we obtain correspond to the most conservative estimate of the bias. Therefore, the posterior shift could well be larger in reality.
For PFS, the shift in $f_w$ goes up to $1\,\sigma_{f_w}$, again determined by the chosen fiducial value of $b_\delta^r$. This shows that neglecting the relative-mode bias contributions can lead to substantial mis-estimates of the WDM mass fraction $f_w$ for current surveys, so that these contributions are essential in real analyses. As expected, the significance of the parameter shift decreases for higher masses, as the relative modes become less significant.

\begin{figure}
    \centering
    \includegraphics[width=\linewidth]{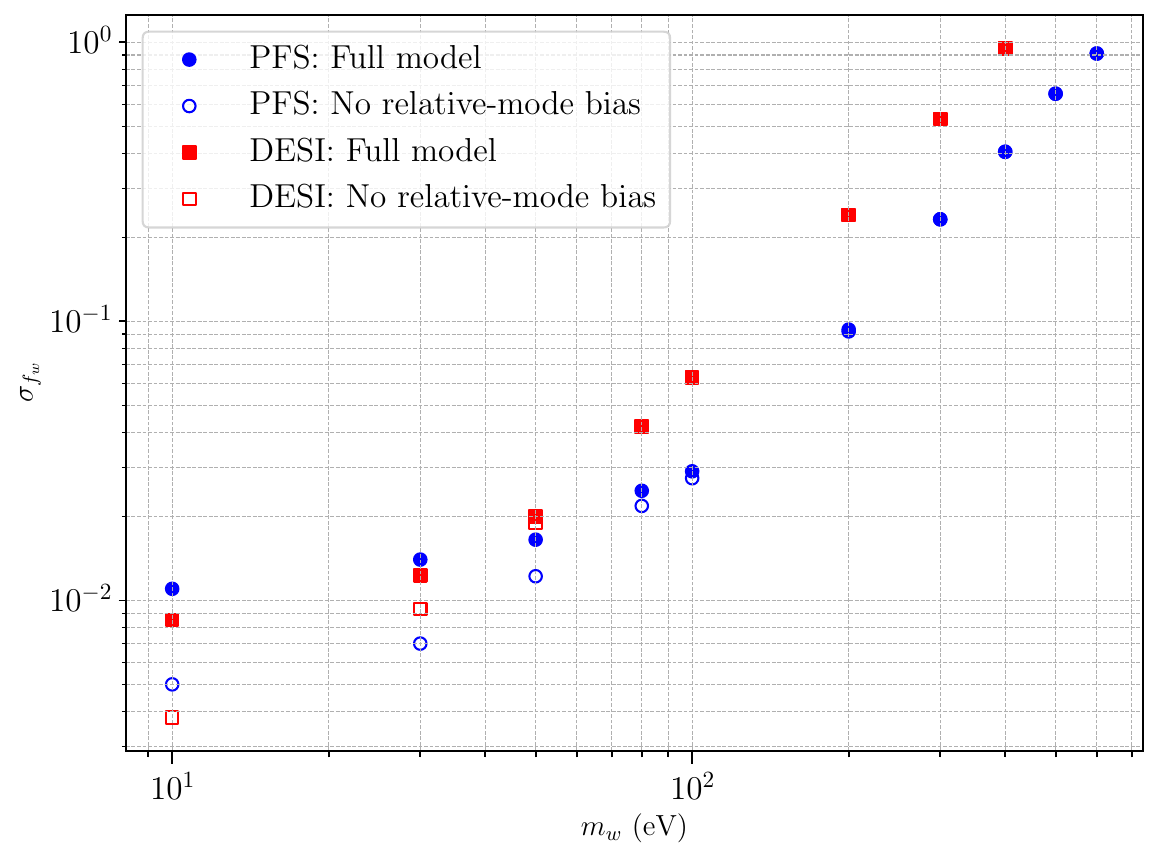}
    \caption{Forecasted constraints on $\sigma_{f_w}$ for warm components with different masses obtained for two different surveys in redshift space. Blue circles (filled and empty) show the results obtained for PFS. The survey specifications, redshift bin, and value of $k_{\mathrm{max}}$ are identical to those used in the analysis shown in Fig.~\ref{fig:corner 10ev matteronly}. Red squares (filled and empty) show the results for DESI. Empty points correspond to a model that neglects contributions from the relative modes, while filled points use the full model with $b_r^{\mathrm{fid}} = f_w^{\mathrm{fid}} b_1^{\mathrm{fid}}$. We used the set of fiducial values $f_w^{\rm fid} = 0.15 ,\, 0.18 ,\, 0.22, \, 0.25, \, 0.30, \, 0.40, \, 0.50, \, 0.60, \, 0.70, \, 0.80 $ respectively for the masses presented in the plot.}
    \label{fig:mass_vs_fraction_combined}
\end{figure}

So far, we have considered constraints at a fixed given WDM mass $m_w$. In an actual analysis, one would derive constraints on the mass fraction $f_w$ while scanning over $m_w$. Fig.~\ref{fig:mass_vs_fraction_combined} shows the expected constraints, $\sigma_{f_w}$, as a function of $m_w$ for the two different surveys, PFS and DESI.
As expected, the constraints monotonically worsen as the mass is increased from the minimum considered here, $m_w = 10\,\mathrm{eV}$. The mass fraction is no longer effectively constrained for $m_w \gtrsim 600\,\mathrm{eV}$ in the case of PFS, and for $m_w \gtrsim 400\,\mathrm{eV}$ for DESI. Although the comoving particle horizon scale $k_\mathrm{ph}$ does not change significantly between redshifts 2 and 1, the nonlinear scale $k_\mathrm{nl}$---and therefore the $k_\mathrm{max}$ that we use in our analysis---decreases at lower redshifts. This limits the ability of lower-redshift surveys to probe high-mass warm components, whose effects manifest on smaller scales. On the other hand, for lighter warm components ($m_w \lesssim 40\,\mathrm{eV}$), DESI yields tighter constraints than PFS. This is mainly because DESI probes lower redshifts, where the clustering amplitude is larger due to the additional growth of structure, increasing the signal in $P_{gg}$. At the same time, in our forecasts (for the adopted fiducial values) the stochastic noise term $P_\epsilon$ is taken to be the same for DESI and PFS, so the measurement is more signal-dominated in DESI. Moreover, DESI also covers a larger survey volume, which further reduces the statistical errors and boosts the overall signal-to-noise, leading to improved constraints on $f_w$ in this mass range.

In both surveys, we observe weakened constraints on $f_w$ for low-mass particles when bias with respect to relative modes is included (full vs open symbols) in Fig.~\ref{fig:mass_vs_fraction_combined}. However, this trend persists up to $m_w \approx 150\,\mathrm{eV}$ for PFS, while it disappears at $m_w \approx 80\,\mathrm{eV}$ for DESI. This difference arises because relative modes become increasingly suppressed at late times, making them less relevant for low-redshift surveys like DESI. In light of our Fisher bias results, we expect systematic biases in $f_w$ to occur over the mass range in Fig.~\ref{fig:mass_vs_fraction_combined} where the relative-mode bias has an impact on constraints. Thus, in an actual analysis, these relative modes should always be included.

\section{Discussion and Conclusions}\label{sec:Conclusion}
In this work, we have derived predictions for large-scale galaxy clustering
in MDM scenarios consisting of both cold and warm components. 
By incorporating relative perturbations between the cold and warm species—--specifically, the relative density and velocity modes—--we extended the standard galaxy bias expansion to account for two-fluid dynamics. Together with \cite{Verdiani:2025jcf}, this represents the first application of a full two-fluid perturbative bias formalism to MDM scenarios. Previous work on related MDM scenarios has neglected the contributions of relative perturbations to galaxy statistics \cite{Xu:2021rwg, Peters:2023asu}.

We derived analytical solutions for the total and relative perturbations using a fluid description, valid outside the free-streaming scale of WDM, $k \ll k_{\rm fs}$ (but with no restriction on the WDM mass fraction $f_w$), and showed that, in addition to a constant relative-density mode $R_0$, a logarithmically growing relative mode, $R_p$, emerges due to the effective pressure of warm dark matter. These modes imprint additional contributions to the galaxy power spectrum, characterized by new bias parameters $b_\delta^r$ and $b_\theta^r$. We incorporated these effects into both real- and redshift-space power spectra, enabling a consistent comparison with future observational data on large scales.

Using a Fisher matrix analysis, we forecast the constraining power of upcoming galaxy surveys such as PFS and DESI on the warm dark matter fraction $f_w$ for a range of particle masses. We found that including relative modes introduces parameter degeneracies—--particularly between $f_w$ and $b_\delta^r$—--which degrade constraints on $f_w$ by up to a factor of 2.5 for light warm components (e.g., $m_w = 10\,\mathrm{eV}$). However, for heavier warm components ($m_w \gtrsim 80\,\mathrm{eV}$), the relative modes become suppressed, and their impact on parameter constraints diminishes significantly. We also examined the potential bias in $f_w$ that arises from incorrect assumptions about the relative bias $b_\delta^r$. In the minimum-mass case, these shifts can reach up to $0.7\sigma$ for DESI and $1\sigma$ for PFS, depending on the choice of fiducial values. These results highlight the sensitivity of $f_w$ to the inclusion of relative modes $R_0, R_p$ particularly at low masses; as expected, the parameter bias becomes insignificant at high masses.

Our results show broad consistency with previous constraints in the literature, although our methodology differs in several aspects. In~\cite{Boyarsky_2009}, constraints on MDM models were derived using Lyman-$\alpha$ forest data, which are particularly sensitive to small-scale power suppression induced by warm components. Their analysis demonstrated that even a modest warm fraction can lead to a significant suppression of structure on small scales. However, their treatment was based on a single-fluid approximation, which does not account for relative perturbations between the cold and warm components.
It would thus be interesting to incoporate the relative perturbations in the model for Lyman-$\alpha$ forest data. Moreover, while Ref.~\cite{Boyarsky_2009}
obtained tighter constraints for heavier warm particles, the constraints are weaker for lighter masses ($m_w \lesssim 100\,\mathrm{eV}$), primarily due to the limited sensitivity of Lyman-$\alpha$ forest observables to larger scales, where low-mass WDM leaves its main imprint. In contrast, our analysis based on large-scale galaxy clustering retains sensitivity to lighter warm species. This highlights the complementary nature of different large-scale structure probes.
It is also worth highlighting that DESI and PFS are competitive in different WDM mass ranges (Fig.~\ref{fig:mass_vs_fraction_combined}), providing a nice complementarity between the two surveys. This is due to their different redshift and number densities.

One of the key advantages of the general EFT bias framework adopted here is that it allows for a well-defined expansion at higher order in perturbation theory, and a joint prediction of other observables such as the galaxy bispectrum (three-point function) or cross-correlations with other tracers. We plan to pursue this in upcoming work. Our formalism can likewise be applied to other, non-WDM mixed scenarios (see \cite{Verdiani:2025jcf} for the ALP case). Beyond extending to higher-order statistics, fast non-linear forward models for mixed dark matter developed in \cite{Parimbelli:2021mtp, Dome:2024hzq} could be used to estimate the bias coefficients associated with relative perturbations by seperate universe approach. In practice, one can impose long-wavelength relative density/velocity backgrounds and use these emulation/halo-model tools to measure the response of halo abundance and clustering, thereby calibrating the relative-mode bias terms.

\section*{Acknowledgement}
We thank Julien Lesgourgues and Julia Stadler for help with the \texttt{CLASS} numerics, and Emmanuele Castorina, Diego Redigolo, Ennio Salvioni, and Francesco Verdiani for discussions. \c{S}\c{C} thanks Beatriz Tucci Schiewaldt and Alejandro P\'erez Fern\'andez for their help with the Fisher analysis, and Eiichiro Komatsu, Toshiki Kurita, and Ivana Nikolac for insightful discussions.
We thank the Galileo Galilei Institute for Theoretical Physics for the hospitality and the INFN for partial support during the completion of this work.

\appendix
\section{Details of the \texttt{CLASS} transfer functions}\label{appendix:CLASS}
In this section, we describe the key steps required to extract the relevant modes needed to build our model using \texttt{CLASS}. Once the code is successfully installed, working with mixtures composed of ordinary cold dark matter (CDM) and non-resonantly produced sterile neutrinos becomes straightforward. The first step is to define the cosmological parameters as follows:

\begin{lstlisting}
    cosmo_values = {
    'omega_b': fiducial_values['omega_b'],
    'N_ncdm': 1,
    'm_ncdm': 10, # in units of eV
    'omega_cdm': fiducial_values['omega_d'] * (1 - fiducial_values['f_w']),
    'omega_ncdm': fiducial_values['omega_d'] * fiducial_values['f_w'],
    'A_s': fiducial_values['A_s'],
    'n_s': fiducial_values['n_s'],
    'H0': fiducial_values['H0'],
    'output': 'tCl,mTk,vTk,mPk',
    'P_k_max_1/Mpc': 10.0,
    'z_pk': 1000.0,
    'gauge': 'sync',
    'ncdm_fluid_trigger_tau_over_tau_k': 51,
    'ncdm_fluid_approximation': 3
}
\end{lstlisting}

In our runs we also enable the built-in \texttt{CLASS} ``fluid approximation'' for the NCDM perturbations on sub-horizon scales, controlled by the flags \texttt{ncdm\_fluid\_approximation} and \texttt{ncdm\_fluid\_trigger\_tau\_over\_tau\_k}. The first selects the specific approximation scheme (3 is the default option) used to replace the full Boltzmann hierarchy of the sterile-neutrino perturbations by a reduced fluid system, while the second sets the (conservative) time threshold for when this switch is performed, such that the
fluid treatment is only applied once a mode is safely inside the horizon.

It is also possible to include multiple non-cold relic species by simply increasing the value of \texttt{N\_ncdm}. Detailed information on the implementation of non-cold relics can be found in~\cite{Lesgourgues_2011}. 

After running the code with the desired cosmological parameters, transfer functions for individual components can be accessed using the \texttt{get\_transfer(z)} function. The total matter density and velocity divergence transfer functions are directly provided, but to isolate the relative modes, one must combine the individual transfer functions $T_{\delta_c}, T_{\delta_w}, T_{\theta_c}, T_{\theta_w}$, which correspond to the density and velocity divergence perturbations of the cold and warm components, respectively. For example, the relative velocity divergence transfer function is given by:

\begin{align}
    T_{\theta_r}(\bm{k},a) = T_{\theta_c}(\bm{k},a) - T_{\theta_w}(\bm{k},a) = \frac{R_p(\bm{k})}{\sqrt{a}} + \frac{R_-(\bm{k})}{a} \,.
\end{align}

Since these two modes scale differently with time, one can compute the transfer functions at two different redshifts and solve algebraically to isolate $R_p(\bm{k})$ and $R_-(\bm{k})$.

\texttt{CLASS} also outputs the sound speed $c_s^2(a)$ of the warm component, which is useful for estimating pressure effects. Because $R_p(\bm{k})$ follows the time evolution of the total matter growing mode, with an additional dependence on $k^2$ and $c_{s_i}^2$, one can cross-check the extracted values at different redshifts as a consistency check.

Similarly, the transfer function for the relative density perturbation can be written as:
\begin{align}
    T_{\delta_r}(\bm{k},a) = T_{\delta_c}(\bm{k},a) - T_{\delta_w}(\bm{k},a) = R_0(\bm{k}) - \ln(a)\frac{R_p(\bm{k})}{H_0} + \frac{2R_-(\bm{k})}{\sqrt{a}H_0} \,.
\end{align}
Since $R_p(\bm{k})$ and $R_-(\bm{k})$ have already been determined, one can now easily solve for $R_0(\bm{k})$.

Finally, it is worth mentioning that the total matter sub-leading mode appearing in Eq.~\eqref{eq:total matter solns} sourced by pressure cannot be directly extracted using transfer functions. It is overshadowed by residual radiation contributions, making it practically impossible to isolate from the individual component transfer functions numerically.

\section{Supplementary material for detailed results}\label{appendix:Full results}

\begin{figure}
    \centering
    \includegraphics[width=\textwidth]{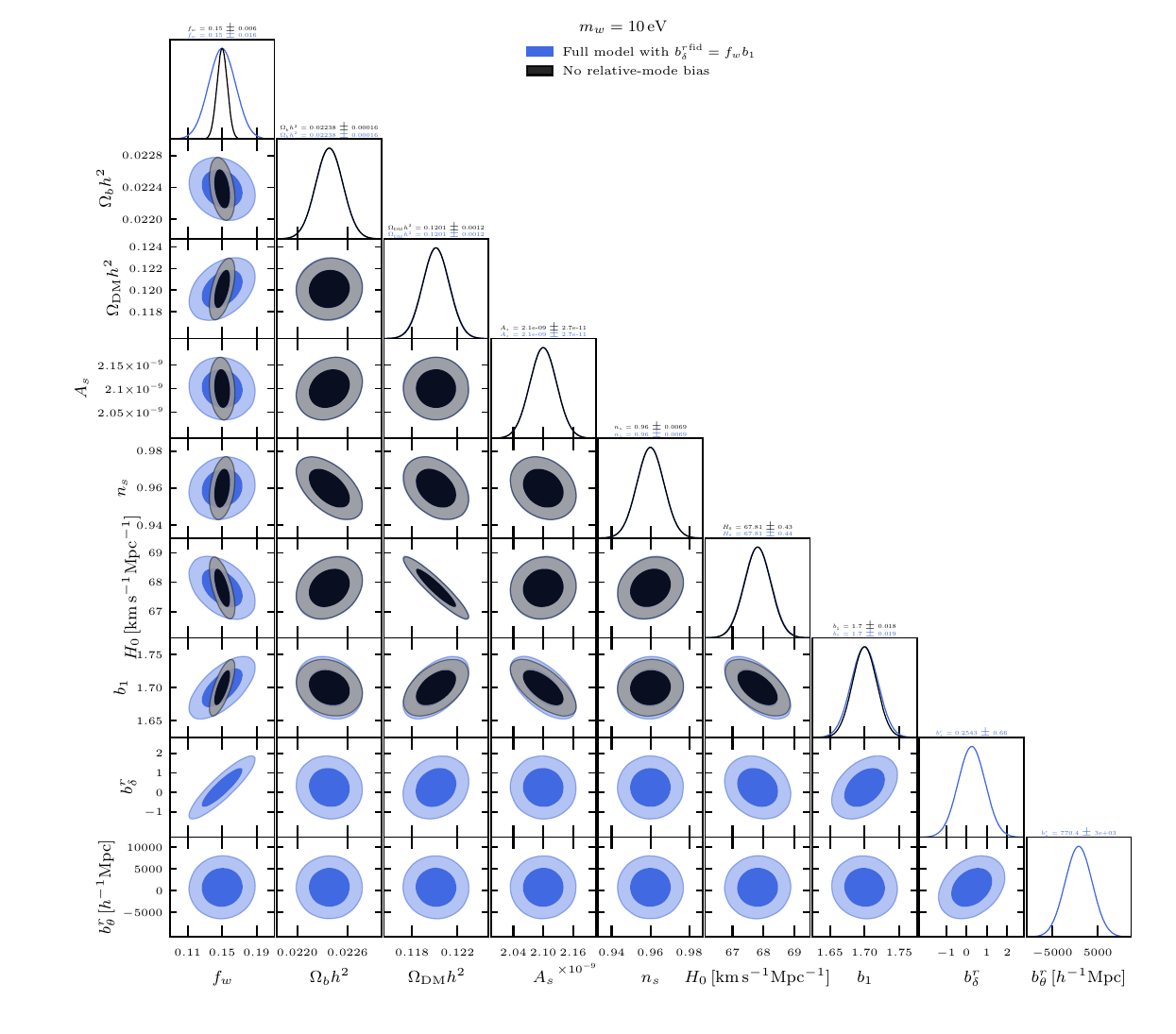}
    \caption{Same as Fig.~\ref{fig:corner 10ev matteronly}, with all cosmological parameters.}
    \label{fig:full real}
\end{figure}

\begin{figure}
    \centering
    \includegraphics[width=\textwidth]{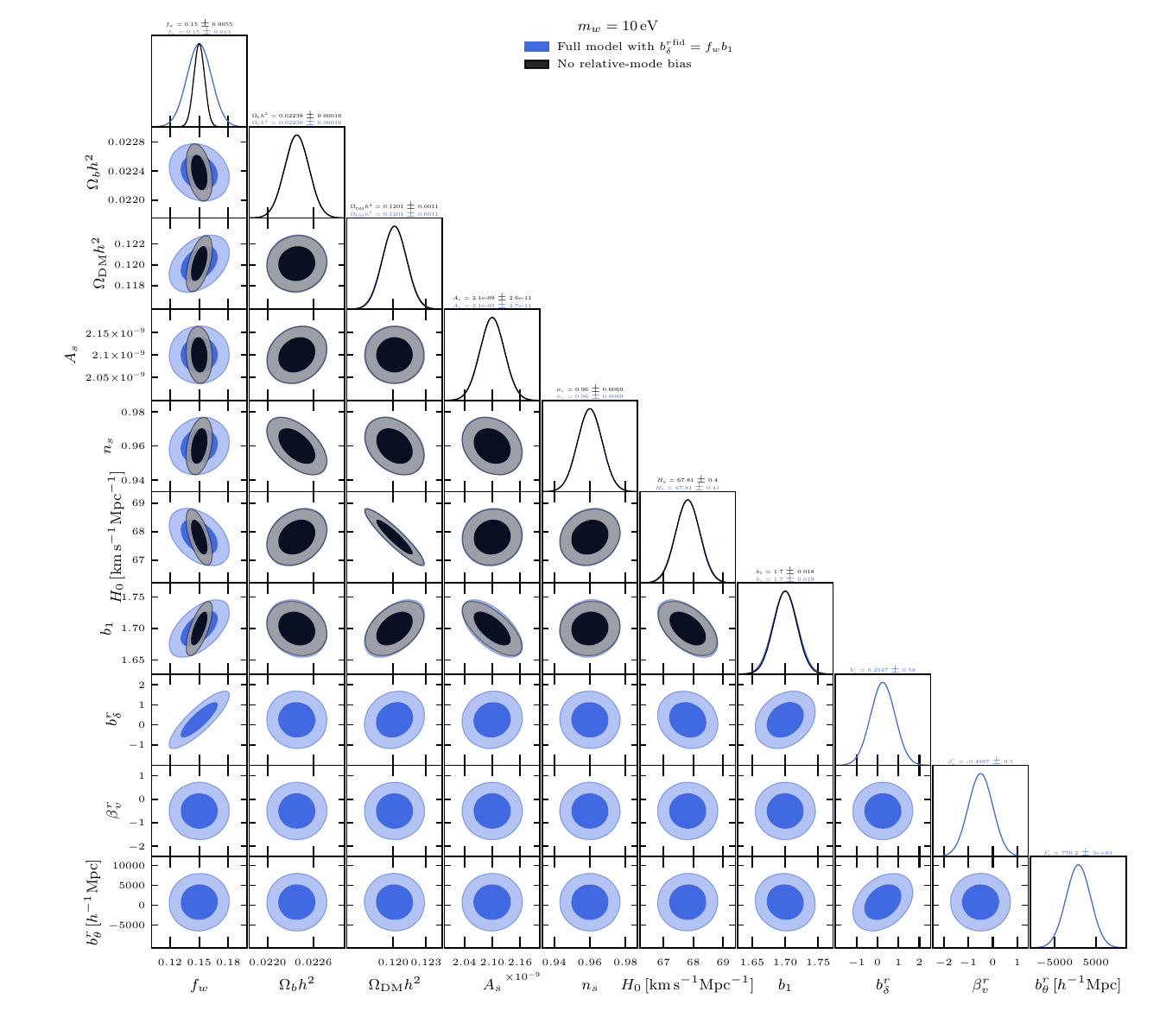}
    \caption{Same as Fig.~\ref{fig:corner 10ev rsd}, with all cosmological parameters.}
    \label{fig:full RSD}
\end{figure}

This section presents the Fisher parameter posteriors for all parameters listed in Eq.~\eqref{eq:parameters} for completeness. We only show the analysis of MDM with $m_w = 10\:\rm eV$, in both real space (\reffig{full real}) and redshift space (\reffig{full RSD}), since this choice is the most informative in terms of parameter degeneracies. As can be seen from these figures, we do not find any significant change in the final constraints or in the structure of the parameter degeneracies for the standard cosmological parameters, whether the relative-mode bias is included or excluded in our model.

\bibliographystyle{ieeetr}
\bibliography{bibliography}

@article{Boyarsky_2009,
   title={Lyman-$\alpha$ constraints on warm and on warm-plus-cold dark matter models},
   volume={2009},
   ISSN={1475-7516},
   url={http://dx.doi.org/10.1088/1475-7516/2009/05/012},
   DOI={10.1088/1475-7516/2009/05/012},
   number={05},
   journal={Journal of Cosmology and Astroparticle Physics},
   publisher={IOP Publishing},
   author={Boyarsky, Alexey and Lesgourgues, Julien and Ruchayskiy, Oleg and Viel, Matteo},
   year={2009},
   month=may, pages={012–012} }

@article{Schmidt_2016,
   title={Effect of relative velocity and density perturbations between baryons and dark matter on the clustering of galaxies},
   volume={94},
   ISSN={2470-0029},
   url={http://dx.doi.org/10.1103/PhysRevD.94.063508},
   DOI={10.1103/physrevd.94.063508},
   number={6},
   journal={Physical Review D},
   publisher={American Physical Society (APS)},
   author={Schmidt, Fabian},
   year={2016},
   month=sep }

@ARTICLE{2009ApJ...700..705S,
       author = {{Shoji}, Masatoshi and {Komatsu}, Eiichiro},
        title = "{Third-Order Perturbation Theory with Nonlinear Pressure}",
      journal = {Astrophys. J.},
         year = 2009,
        month = jul,
       volume = {700},
       number = {1},
        pages = {705-719},
          doi = {10.1088/0004-637X/700/1/705},
archivePrefix = {arXiv},
       eprint = {0903.2669},
 primaryClass = {astro-ph.CO},
       adsurl = {https://ui.adsabs.harvard.edu/abs/2009ApJ...700..705S}
}

@ARTICLE{amin/etal:2025,
       author = {{Amin}, Mustafa A. and {Delos}, M. Sten and {Mirbabayi}, Mehrdad},
        title = "{Structure Formation with Warm White Noise: Effects of Finite Number Density and Velocity Dispersion in Particle and Wave Dark Matter}",
      journal = {arXiv e-prints},
         year = 2025,
        month = mar,
          eid = {arXiv:2503.20881},
        pages = {arXiv:2503.20881},
          doi = {10.48550/arXiv.2503.20881},
archivePrefix = {arXiv},
       eprint = {2503.20881},
 primaryClass = {astro-ph.CO},
       adsurl = {https://ui.adsabs.harvard.edu/abs/2025arXiv250320881A}
}

@ARTICLE{2010PhRvD..82h3520T,
       author = {{Tseliakhovich}, Dmitriy and {Hirata}, Christopher},
        title = "{Relative velocity of dark matter and baryonic fluids and the formation of the first structures}",
      journal = {Phys. Rev. D},
         year = 2010,
        month = oct,
       volume = {82},
       number = {8},
          eid = {083520},
        pages = {083520},
          doi = {10.1103/PhysRevD.82.083520},
archivePrefix = {arXiv},
       eprint = {1005.2416},
 primaryClass = {astro-ph.CO},
       adsurl = {https://ui.adsabs.harvard.edu/abs/2010PhRvD..82h3520T}
}

@ARTICLE{2015JCAP...05..019L,
       author = {{Lewandowski}, Matthew and {Perko}, Ashley and {Senatore}, Leonardo},
        title = "{Analytic prediction of baryonic effects from the EFT of large scale structures}",
      journal = {JCAP},
         year = 2015,
        month = may,
       volume = {2015},
       number = {5},
        pages = {019-019},
          doi = {10.1088/1475-7516/2015/05/019},
archivePrefix = {arXiv},
       eprint = {1412.5049},
 primaryClass = {astro-ph.CO},
       adsurl = {https://ui.adsabs.harvard.edu/abs/2015JCAP...05..019L}
}

@book{Dodelson_Schmidt_2021, place={London, United Kingdom}, edition={2nd}, title={Modern Cosmology}, publisher={Academic Press}, author={Dodelson, Scott and Schmidt, Fabian}, year={2021}}

@article{Lesgourgues_2011,
   title={The Cosmic Linear Anisotropy Solving System (CLASS) IV: efficient implementation of non-cold relics},
   volume={2011},
   ISSN={1475-7516},
   url={http://dx.doi.org/10.1088/1475-7516/2011/09/032},
   DOI={10.1088/1475-7516/2011/09/032},
   number={09},
   journal={Journal of Cosmology and Astroparticle Physics},
   publisher={IOP Publishing},
   author={Lesgourgues, Julien and Tram, Thomas},
   year={2011},
   month=sep, pages={032–032} }

@article{Desjacques_2018,
   title={Large-scale galaxy bias},
   volume={733},
   ISSN={0370-1573},
   url={http://dx.doi.org/10.1016/j.physrep.2017.12.002},
   DOI={10.1016/j.physrep.2017.12.002},
   journal={Physics Reports},
   publisher={Elsevier BV},
   author={Desjacques, Vincent and Jeong, Donghui and Schmidt, Fabian},
   year={2018},
   month=feb, pages={1–193} }

@article{Kaiser,
    author = {Kaiser, Nick},
    title = {Clustering in real space and in redshift space},
    journal = {Monthly Notices of the Royal Astronomical Society},
    volume = {227},
    number = {1},
    pages = {1-21},
    year = {1987},
    month = {07},
    issn = {0035-8711},
    doi = {10.1093/mnras/227.1.1},
    url = {https://doi.org/10.1093/mnras/227.1.1},
    eprint = {https://academic.oup.com/mnras/article-pdf/227/1/1/18522208/mnras227-0001.pdf},
}

@article{PFS,
    author = {Takada, Masahiro and Ellis, Richard S. and Chiba, Masashi and Greene, Jenny E. and Aihara, Hiroaki and Arimoto, Nobuo and Bundy, Kevin and Cohen, Judith and Doré, Olivier and Graves, Genevieve and Gunn, James E. and Heckman, Timothy and Hirata, Christopher M. and Ho, Paul and Kneib, Jean-Paul and Fèvre, Olivier Le and Lin, Lihwai and More, Surhud and Murayama, Hitoshi and Nagao, Tohru and Ouchi, Masami and Seiffert, Michael and Silverman, John D. and Sodré, Laerte, Jr and Spergel, David N. and Strauss, Michael A. and Sugai, Hajime and Suto, Yasushi and Takami, Hideki and Wyse, Rosemary},
    title = {Extragalactic science, cosmology, and Galactic archaeology with the Subaru Prime Focus Spectrograph},
    journal = {Publications of the Astronomical Society of Japan},
    volume = {66},
    number = {1},
    pages = {R1},
    year = {2014},
    month = {02},
    issn = {0004-6264},
    doi = {10.1093/pasj/pst019},
    url = {https://doi.org/10.1093/pasj/pst019},
    eprint = {https://academic.oup.com/pasj/article-pdf/66/1/R1/54684360/pasj\_66\_1\_r1.pdf},
}

@article{Feldman_1994,
   title={Power-spectrum analysis of three-dimensional redshift surveys},
   volume={426},
   ISSN={1538-4357},
   url={http://dx.doi.org/10.1086/174036},
   DOI={10.1086/174036},
   journal={The Astrophysical Journal},
   publisher={American Astronomical Society},
   author={Feldman, Hume A. and Kaiser, Nick and Peacock, John A.},
   year={1994},
   month=may, pages={23} }

@article{Tristram_2024,
   title={Cosmological parameters derived from the final Planck data release (PR4)},
   volume={682},
   ISSN={1432-0746},
   url={http://dx.doi.org/10.1051/0004-6361/202348015},
   DOI={10.1051/0004-6361/202348015},
   journal={Astronomy \& Astrophysics},
   publisher={EDP Sciences},
   author={Tristram, M. and Banday, A. J. and Douspis, M. and Garrido, X. and Górski, K. M. and Henrot-Versillé, S. and Hergt, L. T. and Ilić, S. and Keskitalo, R. and Lagache, G. and Lawrence, C. R. and Partridge, B. and Scott, D.},
   year={2024},
   month=jan, pages={A37} }

@article{2020,
   title={Planck2018 results: VI. Cosmological parameters},
   volume={641},
   ISSN={1432-0746},
   url={http://dx.doi.org/10.1051/0004-6361/201833910},
   DOI={10.1051/0004-6361/201833910},
   journal={Astronomy \& Astrophysics},
   publisher={EDP Sciences},
   author={Aghanim, N. and Akrami, Y. and Ashdown, M. and Aumont, J. and Baccigalupi, C. and Ballardini, M. and Banday, A. J. and Barreiro, R. B. and Bartolo, N. and Basak, S. and Battye, R. and Benabed, K. and Bernard, J.-P. and Bersanelli, M. and Bielewicz, P. and Bock, J. J. and Bond, J. R. and Borrill, J. and Bouchet, F. R. and Boulanger, F. and Bucher, M. and Burigana, C. and Butler, R. C. and Calabrese, E. and Cardoso, J.-F. and Carron, J. and Challinor, A. and Chiang, H. C. and Chluba, J. and Colombo, L. P. L. and Combet, C. and Contreras, D. and Crill, B. P. and Cuttaia, F. and de Bernardis, P. and de Zotti, G. and Delabrouille, J. and Delouis, J.-M. and Di Valentino, E. and Diego, J. M. and Doré, O. and Douspis, M. and Ducout, A. and Dupac, X. and Dusini, S. and Efstathiou, G. and Elsner, F. and Enßlin, T. A. and Eriksen, H. K. and Fantaye, Y. and Farhang, M. and Fergusson, J. and Fernandez-Cobos, R. and Finelli, F. and Forastieri, F. and Frailis, M. and Fraisse, A. A. and Franceschi, E. and Frolov, A. and Galeotta, S. and Galli, S. and Ganga, K. and Génova-Santos, R. T. and Gerbino, M. and Ghosh, T. and González-Nuevo, J. and Górski, K. M. and Gratton, S. and Gruppuso, A. and Gudmundsson, J. E. and Hamann, J. and Handley, W. and Hansen, F. K. and Herranz, D. and Hildebrandt, S. R. and Hivon, E. and Huang, Z. and Jaffe, A. H. and Jones, W. C. and Karakci, A. and Keihänen, E. and Keskitalo, R. and Kiiveri, K. and Kim, J. and Kisner, T. S. and Knox, L. and Krachmalnicoff, N. and Kunz, M. and Kurki-Suonio, H. and Lagache, G. and Lamarre, J.-M. and Lasenby, A. and Lattanzi, M. and Lawrence, C. R. and Le Jeune, M. and Lemos, P. and Lesgourgues, J. and Levrier, F. and Lewis, A. and Liguori, M. and Lilje, P. B. and Lilley, M. and Lindholm, V. and López-Caniego, M. and Lubin, P. M. and Ma, Y.-Z. and Macías-Pérez, J. F. and Maggio, G. and Maino, D. and Mandolesi, N. and Mangilli, A. and Marcos-Caballero, A. and Maris, M. and Martin, P. G. and Martinelli, M. and Martínez-González, E. and Matarrese, S. and Mauri, N. and McEwen, J. D. and Meinhold, P. R. and Melchiorri, A. and Mennella, A. and Migliaccio, M. and Millea, M. and Mitra, S. and Miville-Deschênes, M.-A. and Molinari, D. and Montier, L. and Morgante, G. and Moss, A. and Natoli, P. and Nørgaard-Nielsen, H. U. and Pagano, L. and Paoletti, D. and Partridge, B. and Patanchon, G. and Peiris, H. V. and Perrotta, F. and Pettorino, V. and Piacentini, F. and Polastri, L. and Polenta, G. and Puget, J.-L. and Rachen, J. P. and Reinecke, M. and Remazeilles, M. and Renzi, A. and Rocha, G. and Rosset, C. and Roudier, G. and Rubiño-Martín, J. A. and Ruiz-Granados, B. and Salvati, L. and Sandri, M. and Savelainen, M. and Scott, D. and Shellard, E. P. S. and Sirignano, C. and Sirri, G. and Spencer, L. D. and Sunyaev, R. and Suur-Uski, A.-S. and Tauber, J. A. and Tavagnacco, D. and Tenti, M. and Toffolatti, L. and Tomasi, M. and Trombetti, T. and Valenziano, L. and Valiviita, J. and Van Tent, B. and Vibert, L. and Vielva, P. and Villa, F. and Vittorio, N. and Wandelt, B. D. and Wehus, I. K. and White, M. and White, S. D. M. and Zacchei, A. and Zonca, A.},
   year={2020},
   month=sep, pages={A6} }

@article{shoji2010massive,
  title={Massive neutrinos in cosmology: Analytic solutions and fluid approximation},
  author={Shoji, Masatoshi and Komatsu, Eiichiro},
  journal={Physical Review D—Particles, Fields, Gravitation, and Cosmology},
  volume={81},
  number={12},
  pages={123516},
  year={2010},
  publisher={APS}
}

@misc{lee2025rapidaccuratenumericalevolution,
      title={Rapid and accurate numerical evolution of linear cosmological perturbations with non-cold relics}, 
      author={Nanoom Lee and José Luis Bernal and Sven Günther and Lingyuan Ji and Marc Kamionkowski},
      year={2025},
      eprint={2506.01956},
      archivePrefix={arXiv},
      primaryClass={astro-ph.CO},
      url={https://arxiv.org/abs/2506.01956}, 
}

@article{Bullock_2017,
   title={Small-Scale Challenges to the $\Lambda$CDM Paradigm},
   volume={55},
   ISSN={1545-4282},
   url={http://dx.doi.org/10.1146/annurev-astro-091916-055313},
   DOI={10.1146/annurev-astro-091916-055313},
   number={1},
   journal={Annual Review of Astronomy and Astrophysics},
   publisher={Annual Reviews},
   author={Bullock, James S. and Boylan-Kolchin, Michael},
   year={2017},
   month=aug, pages={343–387} }

@article{
Weinberg2015,
author = {David H. Weinberg  and James S. Bullock  and Fabio Governato  and Rachel Kuzio de Naray  and Annika H. G. Peter },
title = {Cold dark matter: Controversies on small scales},
journal = {Proceedings of the National Academy of Sciences},
volume = {112},
number = {40},
pages = {12249-12255},
year = {2015},
doi = {10.1073/pnas.1308716112},
URL = {https://www.pnas.org/doi/abs/10.1073/pnas.1308716112},
eprint = {https://www.pnas.org/doi/pdf/10.1073/pnas.1308716112}}

@article{Moore1994,
  author       = {Ben Moore},
  title        = {Evidence against dissipation-less dark matter from observations of galaxy haloes},
  journal      = {Nature},
  year         = {1994},
  volume       = {370},
  pages        = {629--631},
  doi          = {10.1038/370629a0},
  url          = {https://doi.org/10.1038/370629a0}
}

@article{deBlok2010,
  author       = {W. J. G. de Blok},
  title        = {The Core-Cusp Problem},
  journal      = {Advances in Astronomy},
  year         = {2010},
  volume       = {2010},
  pages        = {789293},
  doi          = {10.1155/2010/789293},
  url          = {https://doi.org/10.1155/2010/789293}
}

@article{Klypin1999,
  author       = {Anatoly A. Klypin and Andrey V. Kravtsov and Oleg Valenzuela and Francisco Prada},
  title        = {Where Are the Missing Galactic Satellites?},
  journal      = {The Astrophysical Journal},
  year         = {1999},
  volume       = {522},
  number       = {1},
  pages        = {82--92},
  doi          = {10.1086/307643},
  url          = {https://doi.org/10.1086/307643}
}

@article{Moore1999,
  author       = {Ben Moore and Fabio Governato and Thomas Quinn and Joachim Stadel and George Lake},
  title        = {Dark Matter Substructure within Galactic Halos},
  journal      = {The Astrophysical Journal Letters},
  year         = {1999},
  volume       = {524},
  number       = {1},
  pages        = {L19--L22},
  doi          = {10.1086/312287},
  url          = {https://doi.org/10.1086/312287}
}

@article{Asgari_2021,
   title={KiDS-1000 cosmology: Cosmic shear constraints and comparison between two point statistics},
   volume={645},
   ISSN={1432-0746},
   url={http://dx.doi.org/10.1051/0004-6361/202039070},
   DOI={10.1051/0004-6361/202039070},
   journal={Astronomy \& Astrophysics},
   publisher={EDP Sciences},
   author={Asgari, Marika and Lin, Chieh-An and Joachimi, Benjamin and Giblin, Benjamin and Heymans, Catherine and Hildebrandt, Hendrik and Kannawadi, Arun and Stölzner, Benjamin and Tröster, Tilman and van den Busch, Jan Luca and Wright, Angus H. and Bilicki, Maciej and Blake, Chris and de Jong, Jelte and Dvornik, Andrej and Erben, Thomas and Getman, Fedor and Hoekstra, Henk and Köhlinger, Fabian and Kuijken, Konrad and Miller, Lance and Radovich, Mario and Schneider, Peter and Shan, HuanYuan and Valentijn, Edwin},
   year={2021},
   month=jan, pages={A104} }

@article{Abbott2022,
  author       = {{DES Collaboration} and T. M. C. Abbott and F. B. Abdalla and S. Allam and A. Amara and D. Bacon and E. Balbinot and M. Banerji and A. Benoit-Lévy and G. M. Bernstein and E. Bertin and D. Brooks and E. Buckley-Geer and D. L. Burke and M. Carrasco Kind and J. Carretero and F. J. Castander and M. Crocce and C. E. Cunha and L. N. da Costa and J. De Vicente and S. Desai and H. T. Diehl and J. P. Dietrich and P. Doel and T. F. Eifler and A. E. Evrard and A. Fausti Neto and B. Flaugher and P. Fosalba and J. Frieman and J. García-Bellido and E. Gaztanaga and D. W. Gerdes and R. A. Gruendl and K. Honscheid and D. J. James and T. Jeltema and M. D. Johnson and M. W. Gatti and B. Jain and D. Kirk and R. H. Wechsler and J. L. Marshall and P. Martini and C. J. Miller and R. Miquel and J. J. Mohr and E. Neilsen and R. C. Nichol and B. Nord and R. Ogando and A. A. Plazas and A. K. Romer and A. Roodman and E. Sanchez and V. Scarpine and M. Schubnell and I. Sevilla-Noarbe and M. Smith and R. C. Smith and M. Soares-Santos and F. Sobreira and E. Suchyta and G. Tarle and D. Thomas and M. A. Troxel and V. Vikram and A. R. Walker and R. H. Wechsler and J. Zuntz},
  title        = {Dark Energy Survey Year 3 Results: Cosmology from Cosmic Shear and Robustness to Data Calibration},
  journal      = {Physical Review D},
  year         = {2022},
  volume       = {105},
  number       = {2},
  pages        = {023520},
  doi          = {10.1103/PhysRevD.105.023520},
  url          = {https://doi.org/10.1103/PhysRevD.105.023520}
}

@article{Ferreira_2021,
   title={Ultra-light dark matter},
   volume={29},
   ISSN={1432-0754},
   url={http://dx.doi.org/10.1007/s00159-021-00135-6},
   DOI={10.1007/s00159-021-00135-6},
   number={1},
   journal={The Astronomy and Astrophysics Review},
   publisher={Springer Science and Business Media LLC},
   author={Ferreira, Elisa G. M.},
   year={2021},
   month=sep }

@book{KolbTurner1990,
  author    = {Edward W. Kolb and Michael S. Turner},
  title     = {The Early Universe},
  publisher = {Addison-Wesley},
  year      = {1990},
  address   = {Redwood City, CA}
}

@article{DodelsonWidrow1994,
  author  = {Scott Dodelson and Lawrence M. Widrow},
  title   = {Sterile Neutrinos as Dark Matter},
  journal = {Physical Review Letters},
  year    = {1994},
  volume  = {72},
  number  = {1},
  pages   = {17--20},
  doi     = {10.1103/PhysRevLett.72.17},
  url     = {https://doi.org/10.1103/PhysRevLett.72.17}
}

@article{ShiFuller1999,
  author  = {Xun Shi and George M. Fuller},
  title   = {A New Dark Matter Candidate: Nonthermal Sterile Neutrinos},
  journal = {Physical Review Letters},
  year    = {1999},
  volume  = {82},
  number  = {14},
  pages   = {2832--2835},
  doi     = {10.1103/PhysRevLett.82.2832},
  url     = {https://doi.org/10.1103/PhysRevLett.82.2832}
}

@article{Enzi_2021,
   title={Joint constraints on thermal relic dark matter from strong gravitational lensing, the Ly$\alpha$ forest, and Milky Way satellites},
   volume={506},
   ISSN={1365-2966},
   url={http://dx.doi.org/10.1093/mnras/stab1960},
   DOI={10.1093/mnras/stab1960},
   number={4},
   journal={Monthly Notices of the Royal Astronomical Society},
   publisher={Oxford University Press (OUP)},
   author={Enzi, Wolfgang and Murgia, Riccardo and Newton, Oliver and Vegetti, Simona and Frenk, Carlos and Viel, Matteo and Cautun, Marius and Fassnacht, Christopher D and Auger, Matt and Despali, Giulia and McKean, John and Koopmans, Léon V E and Lovell, Mark},
   year={2021},
   month=jul, pages={5848–5862} }

@article{Desjacques:2018pfv,
    author = "Desjacques, Vincent and Jeong, Donghui and Schmidt, Fabian",
    title = "{The Galaxy Power Spectrum and Bispectrum in Redshift Space}",
    eprint = "1806.04015",
    archivePrefix = "arXiv",
    primaryClass = "astro-ph.CO",
    doi = "10.1088/1475-7516/2018/12/035",
    journal = "JCAP",
    volume = "12",
    pages = "035",
    year = "2018"
}

@article{Ivanov:2023yla,
    author = "Ivanov, Mikhail M.",
    title = "{Lyman alpha forest power spectrum in effective field theory}",
    eprint = "2309.10133",
    archivePrefix = "arXiv",
    primaryClass = "astro-ph.CO",
    reportNumber = "MIT-CTP/5607",
    doi = "10.1103/PhysRevD.109.023507",
    journal = "Phys. Rev. D",
    volume = "109",
    number = "2",
    pages = "023507",
    year = "2024"
}

@article{DESI:2016fyo,
    author = "Aghamousa, Amir and others",
    collaboration = "DESI",
    title = "{The DESI Experiment Part I: Science,Targeting, and Survey Design}",
    eprint = "1611.00036",
    archivePrefix = "arXiv",
    primaryClass = "astro-ph.IM",
    reportNumber = "FERMILAB-PUB-16-517-AE",
    month = "10",
    year = "2016"
}

@article{Peters:2023asu,
    author = "Peters, Fabian Hervas and Schneider, Aurel and Bucko, Jozef and Giri, Sambit K. and Parimbelli, Gabriele",
    title = "{Constraining hot dark matter sub-species with weak lensing and the cosmic microwave background radiation}",
    eprint = "2309.03865",
    archivePrefix = "arXiv",
    primaryClass = "astro-ph.CO",
    reportNumber = "NORDITA 2023-032",
    doi = "10.1051/0004-6361/202449195",
    journal = "Astron. Astrophys.",
    volume = "687",
    pages = "A161",
    year = "2024"
}

@article{Xu:2021rwg,
    author = "Xu, Weishuang Linda and Mu{\~n}oz, Julian B. and Dvorkin, Cora",
    title = "{Cosmological constraints on light but massive relics}",
    eprint = "2107.09664",
    archivePrefix = "arXiv",
    primaryClass = "astro-ph.CO",
    doi = "10.1103/PhysRevD.105.095029",
    journal = "Phys. Rev. D",
    volume = "105",
    number = "9",
    pages = "095029",
    year = "2022"
}

@article{Fuss:2022zyt,
    author = "Fu{\ss}, Lea and Garny, Mathias",
    title = "{Decaying Dark Matter and Lyman-{\ensuremath{\alpha}} forest constraints}",
    eprint = "2210.06117",
    archivePrefix = "arXiv",
    primaryClass = "astro-ph.CO",
    reportNumber = "Report-no: TUM-HEP-1420/22",
    doi = "10.1088/1475-7516/2023/10/020",
    journal = "JCAP",
    volume = "10",
    pages = "020",
    year = "2023"
}

@article{Viel:2013fqw,
    author = "Viel, Matteo and Becker, George D. and Bolton, James S. and Haehnelt, Martin G.",
    title = "{Warm dark matter as a solution to the small scale crisis: New constraints from high redshift Lyman-{\ensuremath{\alpha}} forest data}",
    eprint = "1306.2314",
    archivePrefix = "arXiv",
    primaryClass = "astro-ph.CO",
    doi = "10.1103/PhysRevD.88.043502",
    journal = "Phys. Rev. D",
    volume = "88",
    pages = "043502",
    year = "2013"
}

@article{DES:2019ltu,
    author = "Nadler, E. O. and others",
    collaboration = "DES",
    title = "{Milky Way Satellite Census -- II. Galaxy-Halo Connection Constraints Including the Impact of the Large Magellanic Cloud}",
    eprint = "1912.03303",
    archivePrefix = "arXiv",
    primaryClass = "astro-ph.GA",
    reportNumber = "FERMILAB-PUB-19-606-AE",
    doi = "10.3847/1538-4357/ab846a",
    journal = "Astrophys. J.",
    volume = "893",
    pages = "48",
    year = "2020"
}

@article{Barreira:2023uvp,
    author = "Barreira, Alexandre",
    title = "{Constraints on compensated isocurvature perturbations from BOSS DR12 galaxy data}",
    eprint = "2302.01927",
    archivePrefix = "arXiv",
    primaryClass = "astro-ph.CO",
    doi = "10.1088/1475-7516/2023/08/051",
    journal = "JCAP",
    volume = "08",
    pages = "051",
    year = "2023"
}

@article{Senatore:2017hyk,
    author = "Senatore, Leonardo and Zaldarriaga, Matias",
    title = "{The Effective Field Theory of Large-Scale Structure in the presence of Massive Neutrinos}",
    eprint = "1707.04698",
    archivePrefix = "arXiv",
    primaryClass = "astro-ph.CO",
    month = "7",
    year = "2017"
}

@article{Verdiani:2025jcf,
    author = "Verdiani, Francesco and Castorina, Emanuele and Salvioni, Ennio and Sefusatti, Emiliano",
    title = "{The Effective Field Theory of Large Scale Structure for Mixed Dark Matter Scenarios}",
    eprint = "2507.08792",
    archivePrefix = "arXiv",
    primaryClass = "astro-ph.CO",
    month = "7",
    year = "2025"
}

@article{Carrasco:2012cv,
    author = "Carrasco, John Joseph M. and Hertzberg, Mark P. and Senatore, Leonardo",
    title = "{The Effective Field Theory of Cosmological Large Scale Structures}",
    eprint = "1206.2926",
    archivePrefix = "arXiv",
    primaryClass = "astro-ph.CO",
    doi = "10.1007/JHEP09(2012)082",
    journal = "JHEP",
    volume = "09",
    pages = "082",
    year = "2012"
}

@article{Baumann:2010tm,
    author = "Baumann, Daniel and Nicolis, Alberto and Senatore, Leonardo and Zaldarriaga, Matias",
    title = "{Cosmological Non-Linearities as an Effective Fluid}",
    eprint = "1004.2488",
    archivePrefix = "arXiv",
    primaryClass = "astro-ph.CO",
    doi = "10.1088/1475-7516/2012/07/051",
    journal = "JCAP",
    volume = "07",
    pages = "051",
    year = "2012"
}

@inbook{Ivanov:2022mrd,
    author = "Ivanov, Mikhail M.",
    title = "{Effective Field Theory for Large-Scale Structure}",
    eprint = "2212.08488",
    archivePrefix = "arXiv",
    primaryClass = "astro-ph.CO",
    doi = "10.1007/978-981-19-3079-9_5-1",
    year = "2023"
}

@misc{lesgourgues2011cosmiclinearanisotropysolving,
      title={The Cosmic Linear Anisotropy Solving System (CLASS) I: Overview}, 
      author={Julien Lesgourgues},
      year={2011},
      eprint={1104.2932},
      archivePrefix={arXiv},
      primaryClass={astro-ph.IM},
      url={https://arxiv.org/abs/1104.2932}, 
}

@article{Parimbelli:2021mtp,
    author = "Parimbelli, G. and Scelfo, G. and Giri, S. K. and Schneider, A. and Archidiacono, M. and Camera, S. and Viel, M.",
    title = "{Mixed dark matter: matter power spectrum and halo mass function}",
    eprint = "2106.04588",
    archivePrefix = "arXiv",
    primaryClass = "astro-ph.CO",
    doi = "10.1088/1475-7516/2021/12/044",
    journal = "JCAP",
    volume = "12",
    number = "12",
    pages = "044",
    year = "2021"
}

@article{Rampf:2020ety,
    author = "Rampf, Cornelius and Uhlemann, Cora and Hahn, Oliver",
    title = "{Cosmological perturbations for two cold fluids in {\ensuremath{\Lambda}}CDM}",
    eprint = "2008.09123",
    archivePrefix = "arXiv",
    primaryClass = "astro-ph.CO",
    doi = "10.1093/mnras/staa3605",
    journal = "Mon. Not. Roy. Astron. Soc.",
    volume = "503",
    number = "1",
    pages = "406--425",
    year = "2021"
}

@article{Stadler:2018dsa,
    author = "Stadler, Julia and B{\oe}hm, Celine and Mena, Olga",
    title = "{Is it Mixed dark matter or neutrino masses?}",
    eprint = "1807.10034",
    archivePrefix = "arXiv",
    primaryClass = "astro-ph.CO",
    doi = "10.1088/1475-7516/2020/01/039",
    journal = "JCAP",
    volume = "01",
    pages = "039",
    year = "2020"
}

@article{Dome:2024hzq,
    author = {Dome, Tibor and May, Simon and Lagu{\"e}, Alex and Marsh, David J. E. and Johnston, Sarah and Bose, Sownak and Tocher, Alex and Fialkov, Anastasia},
    title = "{Improved halo model calibrations for mixed dark matter models of ultralight axions}",
    eprint = "2409.11469",
    archivePrefix = "arXiv",
    primaryClass = "astro-ph.CO",
    doi = "10.1093/mnras/staf005",
    journal = "Mon. Not. Roy. Astron. Soc.",
    volume = "537",
    number = "1",
    pages = "252--271",
    year = "2025"
}

\end{document}